\documentclass[twocolumn]{aastex631}

\usepackage{chemformula}
\usepackage{multirow}
\usepackage{tabularx}
\usepackage[utf8]{inputenc}
\DeclareUnicodeCharacter{00A0}{\ }
\DeclareSymbolFontAlphabet{\mathrm}    {operators}
\DeclareSymbolFontAlphabet{\mathnormal}{letters}
\DeclareSymbolFontAlphabet{\mathcal}   {symbols}
\DeclareMathAlphabet      {\mathbf}{OT1}{cmr}{bx}{n}
\DeclareMathAlphabet      {\mathsf}{OT1}{cmss}{m}{n}
\DeclareMathAlphabet      {\mathit}{OT1}{cmr}{m}{it}
\DeclareMathAlphabet      {\mathtt}{OT1}{cmtt}{m}{n}
\begin{document}

\title{Thin H$_2$-dominated Atmospheres as Signposts of Magmatic Outgassing on Tidally-Heated Terrestrial Exoplanets}

\author{Rahul Arora}
\affiliation{Lunar and Planetary Laboratory/Department of Planetary Sciences,University of Arizona \\
}

\author{Sukrit Ranjan}
\affiliation{Lunar and Planetary Laboratory/Department of Planetary Sciences,University of Arizona \\
}

\author{Pranabendu Moitra}
\affiliation{Department of Geosciences,University of Arizona \\
}

\author{Ananya Mallik}
\affiliation{Department of Geosciences,University of Arizona \\
}



\begin{abstract}

H$_2$-dominated terrestrial exoplanets are highly accessible to atmospheric characterization via transmission spectroscopy, but such atmospheres are generally thought to be unstable to escape. Here, we propose that close-in, eccentric terrestrial exoplanets can sustain H$_2$-dominated atmospheres due to intense tidally-driven volcanic degassing. We develop an interior–atmosphere framework to assess whether volcanic outgassing can sustain \ch{H2}-dominated atmospheres over geologic timescales ($\geq$1 Gyr). We incorporate interior redox state, tidal heating, volatile inventory, and planetary parameters to compute outgassing fluxes and confront them with energy-limited hydrodynamic escape. We demonstrate that to sustain an H$_2$-dominated atmosphere, a terrestrial exoplanet must have a water-rich basal magma ocean and reduced melts, in addition to high eccentricity. We additionally demonstrate that detection of a specifically thin H$_2$-dominated atmosphere is a sign of current magmatic outgassing. We delineate an ``outgassing zone" (OZ) most favorable to the existence of such planets, and identify the most observationally compelling targets. We propose combining precise mass-radius-eccentricity measurements with JWST constraints on atmospheric mean molecular mass $\mu$ to search for thin H$_2$-dominated atmospheres. Inversely, we argue that robust atmospheric non-detections on OZ exoplanets can constrain the planetary interior, including melt redox state, mantle melt fraction and volatile inventory, and tidal heat flux.

\end{abstract}

\section{Introduction}
Detection of volcanic outgassing on terrestrial exoplanets is a key goal of exoplanet science \citep{Henning2018, quick2020,Gressier_2024,Banerjee_2024, belloarufe2025evidencevolcanicatmospheresubearth}. Detection of volcanic outgassing would test the theory that volcanic outgassing is ubiquitous and the source of terrestrial planet secondary atmospheres \citep{2017aeil.book.....C}, and would constrain exoplanet geodynamics, geochemistry, and evolution \citep{ortenzi2020, liggins2, seligman2023,nichollas2025a}. Volcanic outgassing is particularly key in light of the search for life on exoplanets, for which it is a potential source of abiotic ``false positives" for diverse potential biosignature gases \citep{seager2012, wordsworth_2018, Wogan_2020, Bains2022}. Constraining the prevalence and diversity (magnitude, speciation) of exoplanet volcanic outgassing is therefore key to establishing the ``abiotic baseline", i.e. the range of atmospheric compositions possible in the absence of life against which a putative biosignature detection must be assessed \citep{McMahon2024, Constantinou2025}. 

Detection of volcanic outgassing on terrestrial exoplanets is challenging. Modeling shows that Earth-like outgassing on terrestrial exoplanets is undetectable via near-term facilities/techniques like JWST transmission spectroscopy \citep{hu2013, Misra_2015}. Eccentric terrestrial exoplanets anticipated to have enhanced outgassing due to high tidal heating offer hope \citep{quick2020, seligman2023}, but the biggest challenge for JWST observations is the anticipated high mean molecular mass $\mu$ of terrestrial exoplanet atmospheres, which reduces the planetary scale height and consequently the atmospheric spectral signal \citep{miller2009}. Indeed, even in idealized super-Io scenarios, detectability of exoplanet volcanism is most sensitive to $\mu$ \citep{seligman2023}, and this idealized scenario ignores potential complications due to haze formation surface deposition, and stellar activity \citep{hu2013, Ranjan2023c, Rackham2018}. 

Here we propose a novel, observationally tractable diagnostic for magmatic outgassing on terrestrial exoplanets: the presence of a thin, H$_2$-dominated atmosphere (\ch{H2} volume mixing ratio $\chi_{\ch{H2}} >$ 50\%) atmosphere (Section \ref{subsec:thinH2_tidallyheated}). Such atmospheres are generally unstable to escape, but we show that enhanced outgassing on tidally-heated exoplanets can overcome this limit. We show that the major false positive for this diagnostic, remnant nebular atmospheres, can be mitigated by precise mass/radius measurements to confirm that the atmosphere is thin ($\lesssim 10$ bars) (Section \ref{sec:falsepos}). The major weakness of this diagnostic is that thin H$_2$-dominated atmospheres can only be expected for a narrow range of planetary parameter space (Section \ref{sec:cond}). The major advantage of this diagnostic is that thin H$_2$-dominated atmospheres are highly observationally accessible to JWST, and detection of one would strongly constrain the interior properties of the planet below (Section \ref{sec:snr}). We demarcate an ``outgassing zone" most favorable to the existence of thin H$_2$ atmospheres, and propose that they can be identified by a low-precision JWST spectrum (to detect an H$_2$-dominated atmosphere) in conjunction with a ground-based observational campaign to precisely measure eccentricity (to confirm tidal heating), and mass and radius (to constrain the nebular H$_2$ false positive) (Section \ref{sec:OZ}). Crucially, this test can piggyback off standard atmospheric reconnaissance and characterization campaigns conducted on observationally favorable terrestrial exoplanets (e.g., \citealt{Lim2023, Scarsdale2024, Gressier_2024,AdamsRedai2025}) at no extra cost. We show that nondetections of H$_2$-rich atmospheres on eccentric terrestrial parameters constrain their interior properties.

\section{Methods: Outgassing Calculations}
In this section,  we describe the working of our outgassing model, which calculates the emission fluxes of gases released by magmatic degassing. We developed this model from first principles due to the need to span a very large parameter space (Table~\ref{tab:parameterraneg}) which existing open-source codes were not able to resolve, and we therefore explain its workings in detail. For the rest of our calculations we used previously-published codes and formalisms, which are cited in-place. 

\subsection{Speciation}
Our outgassing model follows \citet{gailard_framework, Wogan_2020} and uses the Equilibrium Constant and Mass Balance method introduced in \citet{Holloway+1987+211+234} to solve for the partial pressure of chemical species outgassed per kg of melt. The models accounts for gas-gas and gas-melt equilibrium along with elemental conservation. It uses a COHS chemical network described in Equation \ref{eq:1} - \ref{eq:10}: 

\begin{equation}
    \ch{H2} + \frac{1}{2}\ch{O2} \longrightarrow \ch{H2O}
    \label{eq:1}
\end{equation}
\begin{equation}
    \ch{CO} + \frac{1}{2}\ch{O2} \longrightarrow \ch{CO2}
    \label{eq:2}
\end{equation}
\begin{equation}
    \ch{CO2} + 2\ch{H2O} \longrightarrow \ch{CH4} + 2\ch{O2}
    \label{eq:3}
\end{equation}
\begin{equation}
    \frac{1}{2}\ch{S2} + \ch{O2} \longrightarrow \ch{SO2} 
    \label{eq:4}
\end{equation}
\begin{equation}
    \frac{1}{2}\ch{S2} + \ch{H2O} \longrightarrow \ch{H2S} + \frac{1}{2}\ch{O2}
    \label{eq:5}
\end{equation}
\begin{equation}
    \ch{CO2}_{(fluid)} + \ch{O}^{2-}_{(melt)} \longrightarrow \ch{CO3}^{2-}_{(melt)} 
    \label{eq:6}
\end{equation}
\begin{equation}
    \ch{H2O}_{(fluid)} + \ch{O}^{2-}_{(melt)} \longrightarrow 2\ch{OH}^{-}_{(melt)} 
    \label{eq:7}
\end{equation}
\begin{equation}
    \ch{H2}_{(fluid)} + \longrightarrow \ch{H2}_{(melt)} 
    \label{eq:8}
\end{equation}
\begin{equation}
    \frac{3}{2}\ch{O2}_{(fluid)} + \ch{O}^{2-}_{(melt)} + \frac{1}{2}\ch{S2}_{(fluid)} \longrightarrow \ch{SO4}^{2-}_{(melt)}
    \label{eq:9}
\end{equation}
\begin{equation}
    \ch{S2}^{-}_{(melt)} + \frac{1}{2}\ch{O2}_{(fluid)} \longrightarrow \ch{O}^{2-}_{(melt)} + \frac{1}{2}\ch{S2}_{(fluid)}
    \label{eq:10}
\end{equation}
Our outgassing calculations solve for three processes simultaneously. First, the volatiles in the melt are either dissolved in the melt or exsolved into bubbles \citep{metrich2007}. Therefore the first process is the gas/melt partitioning, which we account for using solubility relations. The solubilities for hydrogen-bearing molecules are incorporated using \ch{H2} (Equation \ref{eq:11}: \citealt{GAILLARD20032427}):
\begin{equation}
    \chi_{\ch{H2}} = \frac{3.4*10^{-7} p_{\ch{H2}}^{1.28} \mu_{melt}}{2.3 \mu_{\ch{H2}}}
    \label{eq:11}
\end{equation}

\noindent and \ch{H2O} (Equation \ref{eq:12}: \citealt{IACONOMARZIANO20121}):

\begin{equation}
    ln(\chi_{\ch{H2O}}) = a_{\ch{H2O}} ln(p_{\ch{H2O}}) + S_{\ch{H2O}}
    \label{eq:12}
\end{equation}
where \(\chi_{\mathrm{\ch{H2}}}\) and \(\chi_{\mathrm{\ch{H2O}}}\) are mole fraction of \ch{H2} and \ch{H2O} dissolved in the magma after degassing respectively, p\(_{\mathrm{\ch{H2}}}\) and p\(_{\mathrm{\ch{H2O}}}\) are partial pressure of \ch{H2} and \ch{H2O} in gas phase respectively, \(\mu_{\mathrm{melt}}\) is the mean molecular weight of the melt (64 amu for Earth: \citealt{Wogan_2020}), \(\mu_{\mathrm{\ch{H2}}}\) is molar mass of \ch{H2} (2.016 amu), S\(_{\mathrm{\ch{H2O}}}\) and a\(_{\mathrm{\ch{H2O}}}\)  are \ch{H2O} solubility and solubility constant respectively. For details, see Appendix \ref{sec:appsolub}.

Carbon is assumed to dissolve as \ch{CO2} (Equation \ref{eq:13}: \citealt{IACONOMARZIANO20121}):
\begin{equation}
    ln(\chi_{\ch{CO2}}) = \chi_{\ch{H2O}} d_{\ch{H2O}} + a_{\ch{CO2}} ln(p_{\ch{CO2}}) + S_{\ch{CO2}}
    \label{eq:13}
\end{equation}
where \(\chi_{\mathrm{\ch{CO2}}}\) and p\(_{\mathrm{\ch{CO2}}}\) are dissolved mole fraction and partial pressure of \ch{CO2} respectively. S\(_{\mathrm{\ch{CO2}}}\) is the \ch{CO2} solubility, and d\(_{\mathrm{\ch{H2O}}}\) and a\(_{\mathrm{\ch{CO2}}}\) are solubility constants respectively. For details, see Appendix \ref{sec:appsolub}.

Sulfur solubility is included in two forms, sulfide (S$^{2-}$) (Equation \ref{eq:14}: \citealt{oneill2002}):
\begin{equation}
    ln(\chi_{S^{2-}}) = \frac{C_{S^{2-}}p_{\ch{S2}}^{0.5}\mu_{melt}}{f_{\ch{O2}}^{0.5}\mu_S}
    \label{eq:14}
\end{equation}

\noindent and sulfate (S$^{6+}$) (Equation \ref{eq:15}: \citealt{oneill2022368}):

\begin{equation}
    ln(\chi_{S^{6+}}) = \frac{C_{S^{6+}}p_{\ch{S2}}^{0.5}f_{\ch{O2}}^{1.5}\mu_{melt}}{\mu_S}
    \label{eq:15}
\end{equation}
\noindent here, $\chi_{\ch{S^{2-}}}$ and $\chi_{\ch{S^{6+}}}$ are the mole fractions of \ch{S^{2-}} and \ch{S^{6+}} in the melt, $p_{\ch{S2}}$ is the partial pressure of \ch{S2} gas, $f_{\ch{O2}}$ is the oxygen fugacity, and $\mu_S$ is the molar mass of sulfur. $C_{\ch{S^{2-}}}$ and $C_{\ch{S^{6+}}}$ are empirically derived solubility constants for the respective sulfur species. The contrasting dependencies on $f_{\ch{O2}}$ in the two expressions appearing as $f_{\ch{O2}}^{-0.5}$ for reduced sulfur and $f_{\ch{O2}}^{1.5}$ for oxidized sulfur shows the redox state sensitivity on partitioning behavior of sulfur in magmatic systems, with \ch{S^{6+}} dissolution favored under oxidizing conditions and \ch{S^{2-}} dissolution under reducing ones.\\

\noindent Once these volatiles are separated into melt and gas phases, the gas phase molecules present in the bubbles will mutually react to thermochemical equilibrium. Hence, the second process to solve for is the gas-gas equilibrium, which are represented by Equation \ref{eq:16}-\ref{eq:20}: 

\begin{equation}
    K_1 = \frac{p_{\ch{H2O}}}{p_{\ch{H2}}*f_{\ch{O2}}^{0.5}}
    \label{eq:16}
\end{equation}
\begin{equation}
    K_2 = \frac{p_{\ch{CO2}}}{p_{CO}*f_{\ch{O2}}^{0.5}}
    \label{eq:17}
\end{equation}
\begin{equation}
    K_3 = \frac{p_{CH_4}*f_{\ch{O2}}^2}{p_{\ch{H2O}}^2*p_{\ch{CO2}}}
    \label{eq:18}
\end{equation}
\begin{equation}
    K_4 = \frac{p_{\ch{SO2}}}{p_{\ch{S2}}^{0.5}*f_{\ch{O2}}}
    \label{eq:19}
\end{equation}
\begin{equation}
    K_5 = \frac{p_{\ch{H2}S}*f_{\ch{O2}}^{0.5}}{p_{\ch{H2O}}*p_{\ch{S2}}^{0.5}}
    \label{eq:20}
\end{equation}

We obtain the temperature dependent fits for $K_1-K_5$ from \citet{symonds1993}. Additionally, we assume a closed system following \citet{Wogan_2020}, dictating elemental conservation (Equation \ref{eq:21}-\ref{eq:23}): 

\begin{equation}
    \begin{aligned}
    \frac{m_{\ch{CO2}_{tot}}*\mu_{melt}}{\mu_{\ch{CO2}}} = \frac{(p_{\ch{CO2}}+p_{\ch{CO}}+p_{\ch{CH4}})*\alpha_{gas}}{P} \\ +(1-\alpha_{gas})*\chi_{\ch{CO2}}
    \label{eq:21}
   \end{aligned}
\end{equation}
\begin{equation}
    \begin{aligned}
    \frac{m_{\ch{H2O}_{tot}}*\mu_{melt}}{\mu_{\ch{H2O}}} = \frac{(p_{\ch{H2}}+p_{\ch{H2O}}+2*p_{\ch{CH4}})*\alpha_{gas}}{P} \\+(1-\alpha_{gas})*\chi_{\ch{H2O}}
    \label{eq:22}
    \end{aligned}
\end{equation}
\begin{equation}
\begin{aligned}
    \frac{m_{S_{tot}}*\mu_{melt}}{\mu_{S}} = \frac{(p_{\ch{SO2}}+p_{\ch{H2S}}+2*p_{\ch{S2}})*\alpha_{gas}}{P} \\+(1-\alpha_{gas})*\chi_{S}
    \label{eq:23}
    \end{aligned}
\end{equation}
where $m_{\ch{H2O}_{tot}}$, $m_{\ch{CO2}_{tot}}$, $m_{\ch{S}_{tot}}$ is the initial mass fraction of \ch{H2O}, \ch{CO2} and \ch{S} in undegassed melt respectively and $\alpha_{gas}$ is the mole fraction of gas in the gas+melt system 
The oxygen concentration is set by the melt \textit{f}\ch{O2}, which we prescribe in reference to the fayalite-magnetite-quartz (FMQ) buffer \citep{frost1991}:

\begin{equation}
    \begin{aligned}
    log_{10}(f_{\ch{O2}}) = A - (B/T) +  \frac{C*(P-1.0)}{T} \\+ \Delta FMQ
    \label{eq:24}
    \end{aligned}
\end{equation}
where P is the total pressure, T is the Temperature, A = 8.375, B =-25096.3 and C = 0.11 \citep{frost1991}.

\subsection{Graphite Saturation}
\label{sec:graph}
In strongly reducing magmatic environments, carbon is expected to be stored primarily in the form of graphite \citep{dasgupta_2013,ortenzi2020,thompson2022}. Under such conditions, the abundance of outgassed \ch{CO2} is significantly limited by the solubility of carbonate ions (\ch{CO3^{2-}}) in the melt, which is in turn governed by the ambient oxygen fugacity ($f_{\mathrm{O}_2}$). This graphite saturation acts as a natural sink for carbon, and suppresses carbon outgassing \citep{dasgupta_2013,ortenzi2020}, influencing the redox-dependent speciation in our outgassing model.

To incorporate this suppression mechanism, we calculate the equilibrium concentration of carbonate ions in the melt following \citet{ortenzi2020}. The mole fraction of carbonate ions in the melt ($X^{\text{melt}}_{\ch{CO_3^{2-}}}$) is given by Equation \ref{eq:25}:

\begin{equation}
X^{\text{melt}}_{\ch{CO_3^{2-}}} = \frac{K_a K_b f_{\ch{O}_2}}{1 + K_1 K_2 f_{\ch{O}_2}}
\label{eq:25}
\end{equation}

where $K_a$ and $K_b$ are temperature- and pressure-dependent equilibrium constants. These are evaluated using the expressions Equation \ref{eq:26}, \ref{eq:27}\citep{ortenzi2020}:
\begin{equation}
\begin{aligned}
\log_{10} K_a = 40.07639 - 2.53932 \times 10^{-2} T \\+5.27096\times 10^{-6} T^2 
+ \frac{0.0267 (p - 1)}{T} 
\label{eq:26}
\end{aligned}
\end{equation}

\begin{equation}
\begin{aligned}
\log_{10} K_b = -6.24763 - \frac{282.56}{T} - \frac{0.119242 (P - 1000)}{T}
\label{eq:27}
\end{aligned}
\end{equation}

\noindent where $T$ is temperature in Kelvin and $P$ is the total pressure in bar.

The total mole fraction of \ch{CO2} dissolved in the melt ($X^{\text{melt}}_{\ch{CO_2}}$) is then computed from the equilibrium carbonate content as (Equation \ref{eq:28}; \citealt{ortenzi2020}):

\begin{equation}
X^{\text{melt}}_{\ch{CO_2}} = \left[\frac{M_{\ch{CO_2}}}{\mathrm{fwm}} X^{\text{melt}}_{\ch{CO_3^{2-}}}\right]\left/\left[1 - \left(1 - \frac{M_{\ch{CO_2}}}{\mathrm{fwm}}\right) X^{\text{melt}}_{\ch{CO_3^{2-}}}\right]\right.
\label{eq:28}
\end{equation}

where $M_{\ch{CO_2}}$ is the molar mass of \ch{CO_2} and fwm is the formula weight of the melt per oxygen atom (taken to be 36.594, consistent with 1921 Kilauea tholeiitic basalt composition) following the approch of \cite{holloway1992}.

By explicitly calculating $X^{\text{melt}}_{\ch{CO_2}}$ under varying oxygen fugacity scenarios, our model captures the suppression of \ch{CO2} in reducing interiors and improves the realism of predicted outgassing compositions \citep{ortenzi2020,thompson2022}. For this work, we integrate this formulation directly into our self-developed outgassing code to evaluate the resulting surface fluxes of \ch{CO2} and its role as an oxygen source.

\subsection{Outgassing Flux}

So far, we have calculated the partial pressure of chemical species exsolved per kg of melt. To convert this into the global-average outgassing flux we follow \citet{Wogan_2020} to calculate outgassing rate in mol/yr and then convert to flux (molecules cm$^{-2}$ s$^{-1}$) (Equation \ref{eq:outgflux})  :

\begin{equation}
    \phi_{X, P} = \frac{\alpha_{\mathrm{gas}}*N_A}{\mu_{\mathrm{magma}}(1-\alpha_{\mathrm{gas})}(4\pi R_{p}^2)}\frac{p_{X}}{P_{\mathrm{tot}}} Q
    \label{eq:outgflux}
\end{equation}

\noindent
Here, \( \phi_{X, P} \) is the outgassing flux of X species, $R_p$ is the radius of the planet, $p_{X`}$ is the partial pressure of outgassed species X, $P_{\mathrm{tot}}$ is the total surface pressure, and $Q$ is the magma production rate of the planet. 

We estimate $Q$ under the simplifying assumption that the magma production rate scales linearly with the interior heat flux of the planet (e.g., \citealt{ Hu2023, Gkouvelis2025}):
\begin{align}
    Q=Q_{\oplus}\frac{\eta_{og}\dot{E}_{\mathrm{Heat}, P}}{\eta_{og,\oplus}\dot{E}_{\mathrm{Heat}, \oplus}}
    \label{eq:meltpro}
\end{align}
\noindent where \( \dot{E}_{\mathrm{Heat}, P} \) is the planet’s total interior rate flux and the Earth heat rate and magma production rate are are Q$_{\oplus}$ = 25 km$^3$/yr and \( \dot{E}_{\mathrm{Heat}, \oplus} = 4.7 \times 10^{20}\,\mathrm{erg\,s^{-1}} \) \citep{holland2002, Turcotte_Schubert_2014}. \( \eta_{\mathrm{og}} \) is the outgassing efficiency, i.e. the fraction of interior heat that is transported advectively as magma. We assume a baseline \( \eta_{\mathrm{og}} =\eta_{\mathrm{og, \oplus}}= 0.1 \), corresponding to Earth-like outgassing \citep{Hu2023}. However, this efficiency can vary significantly across planets depending on mantle composition, volatile content, and lithospheric structure \citep{guimond2021}. For example, on Io, a larger fraction of interior heat is dissipated by advective transport (which produces melt and therefore outgassing) compared to Earth, implying \( \eta_{\mathrm{og}} = 1 \) \citep{Moore2007, Hu2023}. We consequently explore \( \eta_{\mathrm{og}} = 0.01-1 \), embracing the range in the solar system and considering the possibility of low outgassing efficiency as well. Additional variation in \( \eta_{\mathrm{og}} \) may be possible due to variations in the intrusive-to-extrusive outgassing ratio, but as we are not aware of variation within the Solar System from the Earth-like value ($\sim0.1$; \citealt{Hu2023}), we do not consider this source of variability. 

To estimate \( \dot{E}_{\mathrm{Heat}, P} \), we consider intensely tidally heated planets where the interior heat flux is dominated by tidal heating, and follow \citet{seligman2023}:

\begin{equation}
\begin{aligned}
\dot{E}_{\mathrm{Heat}} =  \eta_{\mathrm{Heat}}&\left(3.4 \times 10^{25} \,\mathrm{erg\,s^{-1}}\right) \left( \frac{P}{1\,\mathrm{d}} \right)^{-5} \left( \frac{R_P}{R_\oplus} \right)^5 \\
& \times \left( \frac{e}{10^{-2}} \right)^2 \left( \frac{\Im(\tilde{k}_2)}{10^{-2}} \right)
\end{aligned}
\label{eq:tidalheat}
\end{equation}

\noindent
where \( P \) is the orbital period (in days), \( e \) is orbital eccentricity, and \( \Im(\tilde{k}_2) \) is the imaginary part of the complex Love number. Note that this approach may overestimate total tidal heating if the runaway melting regime of \citet{Peale1979} is not accessed; this estimate therefore corresponds to an upper limit on $\dot{E}_{\mathrm{Heat}}$ \citep{nichollas2025a}. For this reason, we introduce a heating efficiency parameter $\eta_{\mathrm{Heat}}$, which is allowed to range from $\eta_{\mathrm{Heat}}=0.01-1$, with the upper limit corresponding to the runaway melting limiting case assumed by \citet{seligman2023}, while the lower limit corresponds to the suppressed heating rate proposed by \citet{nichollas2025a}.
\section{Thin low-$\mu$ Atmospheres as Volcanism Diagnostics}

We propose a thin \ch{H2}-dominated ($\chi_{\ch{H2}} >$ 50\%) atmosphere as a diagnostic for magmatic outgassing on rocky exoplanets. Such atmospheres are generally unstable to atmospheric escape, but we argue that enhanced outgassing on tidally-heated exoplanets can overcome this limit. The major advantage of this diagnostic is that it is highly accessible to observational characterization with JWST, because the low $\mu$ implied by a \ch{H2}-dominated atmosphere will render atmospheric spectral features readily detectable \citep{miller2009,Burrows_2014}. The major weakness of this diagnostic is that such atmospheres can only be expected for a narrow range of geochemical parameter space. The flip side of this weakness is that detection of a thin \ch{H2}-dominated atmosphere strongly constrains the geochemistry of the underlying planet. We consider LP 98-59d as a potential case study of such an atmosphere \citep{Gressier_2024, Banerjee_2024}.

\subsection{Thin \ch{H2}-dominated Atmospheres Require Replenishment}
\label{sec:lowmmmw}

Thin \ch{H2}-dominated atmospheres on rocky planets are generally unstable to escape because \ch{H2} is light and readily escapes from their weak gravitational fields \citep{Hu2023}.  Instead, rocky planets are expected to accumulate volcanogenic heavy volatiles like \ch{N2} and \ch{CO2} in secondary atmospheres \citep{liggins2}. The rocky planets of the solar system (Earth, Mars, Venus) serves as an examples of this paradigm, having lost their natal \ch{H2} atmospheres in favor of high-$\mu$ secondary atmospheres.

We illustrate the general instability of \ch{H2}-dominated atmospheres using the energy-limited escape approximation following \citet{Hu2023} and converted from mass loss rate (g/s) to number escape fluxmolecule cm$^{-2}$ s${-1}$ (Equation \ref{eq:els}):
\begin{equation}
    \phi_{esc} = \frac{\pi \eta F_{XUV}R^3_Pa_r^2}{GM_PK} \frac{N_A}{\mu_{\ch{H2}}(4\pi R_P^2)}
    \label{eq:els}
\end{equation}
\noindent where F$_{XUV}$ (5-1240 \AA{}; \citealt{Hu2023}) is the sum of X-ray and EUV flux received by the planet \citep{Tian2015,Gronoff_2020,Hu2023}, R$_P$ is the radius of the planet, M$_P$ is the mass of the planet, $a_r$ is the ratio between the X-ray/EUV absorbing radius and the (optical) planetary radius, K is a factor that accounts for the Roche lobe effect \citep{Erkaev2017} and $\eta$ is the efficiency of the escape. We adopt $\eta$ = 0.1, a = 1, and K = 1 as a conservative estimate of the escape rate for our calculations following \cite{Hu2023}. We find that for diverse exoplanets, the lifetime of a 10-bar H$_2$ atmosphere is geologically short ($<<1$ Gyr; Table~\ref{tab:life}). Therefore thin H$_2$-dominated atmospheres on planets orbiting mature stars ($\geq1$ Gyr) imply active \ch{H2} outgassing at high rates to counterbalance escape. 
\begin{table*}[]
\hspace*{-6em}
\begin{tabular}{llllllll}
\hline
Planet       & R$_P$        & M$_P$        & a      & F$_{\mathrm{XUV}}$      & Escape rate    & Lifetime (Myrs) & References                                              \\ \hline
             & R$_{\oplus}$ & M$_{\oplus}$ & AU     & ergs s$^{-1}$ cm$^{-2}$ & H$_2$ bars/Myr & No outgasing    &                                                         \\ \hline
Earth \tablenotemark{*}    & 1.00         & 1.00         & 1.00   & 3.88                    & 0.00479        & 2090            & \citep{ribas2005}                          \\
Venus \tablenotemark{*}    & 0.94         & 0.815        & 0.72   & 7.48                    & 0.00983        & 1020            & \citep{ribas2005}                          \\
Mars \tablenotemark{*}    & 0.53         & 0.107        & 1.52   & 1.68                    & 0.00391        & 2560            & \citep{ribas2005}                          \\
TRAPPIST-1 b & 1.12         & 1.02         & 0.0115 & 2800                    & 3.09           & 3.24            & \citep{Grimm_2018,Becker_2020}          \\
TRAPPIST-1 c & 1.09         & 1.16         & 0.0158 & 1500                    & 1.69           & 5.93            & \citep{Grimm_2018,Becker_2020}          \\
TRAPPIST-1 d & 0.784        & 0.297        & 0.0223 & 753                     & 1.19           & 8.43            & \citep{Grimm_2018,Becker_2020}          \\
TRAPPIST-1 e & 0.91         & 0.772        & 0.0293 & 436                     & 0.592          & 16.9            & \citep{Grimm_2018,Becker_2020}          \\
TRAPPIST-1 f & 1.05         & 0.934        & 0.0385 & 252                     & 0.297          & 33.6            & \citep{Grimm_2018,Becker_2020}          \\
TRAPPIST-1 g & 1.15         & 1.15         & 0.0469 & 170                     & 0.183          & 54.6            & \citep{Grimm_2018,Becker_2020}          \\
TRAPPIST-1 h & 0.773        & 0.331        & 0.0619 & 97.5                    & 0.156          & 64.2            & \citep{Grimm_2018,Becker_2020}          \\
L 98-59b     & 0.836        & 0.46         & 0.0223 & 1110                    & 1.63           & 6.12            & \citep{Cadieux2025,muscles_extension} \\
L 98-59c     & 1.33         & 2.00         & 0.0386 & 369                     & 0.343          & 29.2            & \citep{Cadieux2025,muscles_extension} \\
L 98-59d     & 1.63         & 1.64         & 0.0494 & 225                     & 0.171          & 58.5            & \citep{Cadieux2025,muscles_extension} \\
LP 791-18 d  & 1.03         & 0.90         & 0.0190 & 8010                    & 9.58           & 1.04            & \citep{peterson2023,muscles_extension}    \\ \hline
\end{tabular}

\tablenotetext{*}{\href{https://nssdc.gsfc.nasa.gov/planetary/factsheet/}{https://nssdc.gsfc.nasa.gov/planetary/factsheet/}}
\caption{Planetary properties and atmospheric escape estimates. Column 1 lists the planet names, followed by their radius in Earth radii (Column 2) and mass in Earth masses (Column 3). Column 4 provides the incident XUV flux in ergs s$^{-1}$/cm$^{2}$. Column 5 presents the calculated atmospheric escape rate in  \ch{H2} bars per million years (Myr), and Column 6 shows the estimated lifetime of a 10-bar \ch{H2} atmosphere under the given escape conditions.} 
\label{tab:life}
\end{table*}
\subsection{Volcanic outgassing as a source of \ch{H2}} \label{sec:tidal}
We suggest terrestrial exoplanets subjected to enhanced volcanism due to high tidal heating may outgas \ch{H2} at a high enough rate to counterbalance hydrodynamic escape. Volcanism is the main source of molecular hydrogen to the surface of terrestrial planets, counterbalanced by escape to space \citep{2017aeil.book.....C}. The idea that \ch{H2} outgassing can outpace escape was first articulated by \citet{tian2005} for early Earth, who argued that prior work had overestimated \ch{H2} escape rates by two orders of magnitude and that early Earth might have hosted an \ch{H2}-rich atmosphere ($30\%$ v/v). However, the escape calculations of \citet{tian2005} were later criticized on methodological grounds \citep{Kulikov2007}. There is now general consensus that Earth-like outgassing cannot sustain an \ch{H2}-dominated atmosphere against escape, particularly for planets orbiting XUV-active M-dwarf stars that are the main targets for JWST \citep{Hu2023}.

However, exoplanet volcanism may not be Earth-like. In particular, tightly-orbiting M-dwarf planets in eccentric orbits will experience extreme tidal heating, which is expected to drive volcanism rates much higher than Earth's \citep{jakson2008, Driscoll2015}. \citet{quick2020} and \citet{seligman2023} apply tidal theory to identify rocky exoplanets expected to possess  orders of magnitude higher internal heat flux compared to  modern Earth. These high levels of tidal heating are predicted to lead to increased magma production and outgassing.


We propose that planets subject to extreme tidal heating may outgas enough H$_2$ to counterbalance hydrodynamic atmospheric escape and sustain H$_2$-dominated atmospheres. To quantitatively illustrate this, we calculate outgassing for a subset of observationally favorable terrestrial planets with high nominal eccentricity (Table~\ref{tab:life}), which have been previously highlighted as targets for detecting exoplanet volcanism \citep{peterson2023,seligman2023}. We assume Earth-like outgassing (i.e. the same pH$_2$ per kg of melt), but allow the melt production rate to increase following Equation~\ref{eq:meltpro}. This isolates the effect of the increased interior heat flux from other controlling variables like melt \textit{f}O$_2$ and P$_{\mathrm{tot}}$. We permit the combined outgassing and heating efficiency parameter, \(\eta_{\mathrm{og}}\eta_{\mathrm{Heat}}\), to vary from \(10^{-4}\) to \(1\), spanning the range suggested from interior modeling and the Solar System \citep{Moore2007,Hu2023}. Contrary to past work \citep{liggins1, Hu2023}, we find that for physically plausible \(\eta_{\mathrm{og}}\eta_{\mathrm{Heat}}\), it is possible for Earth-like outgassing to outpace escape, driven by higher outgassing due to orders of magnitude higher tidally-driven heat flux (Figures \ref{fig:orbital}, \ref{fig:inventime}).

\subsection{Outgassing zone}\label{sec:OZ}
\begin{figure}
    \centering
    \includegraphics[scale =0.35]{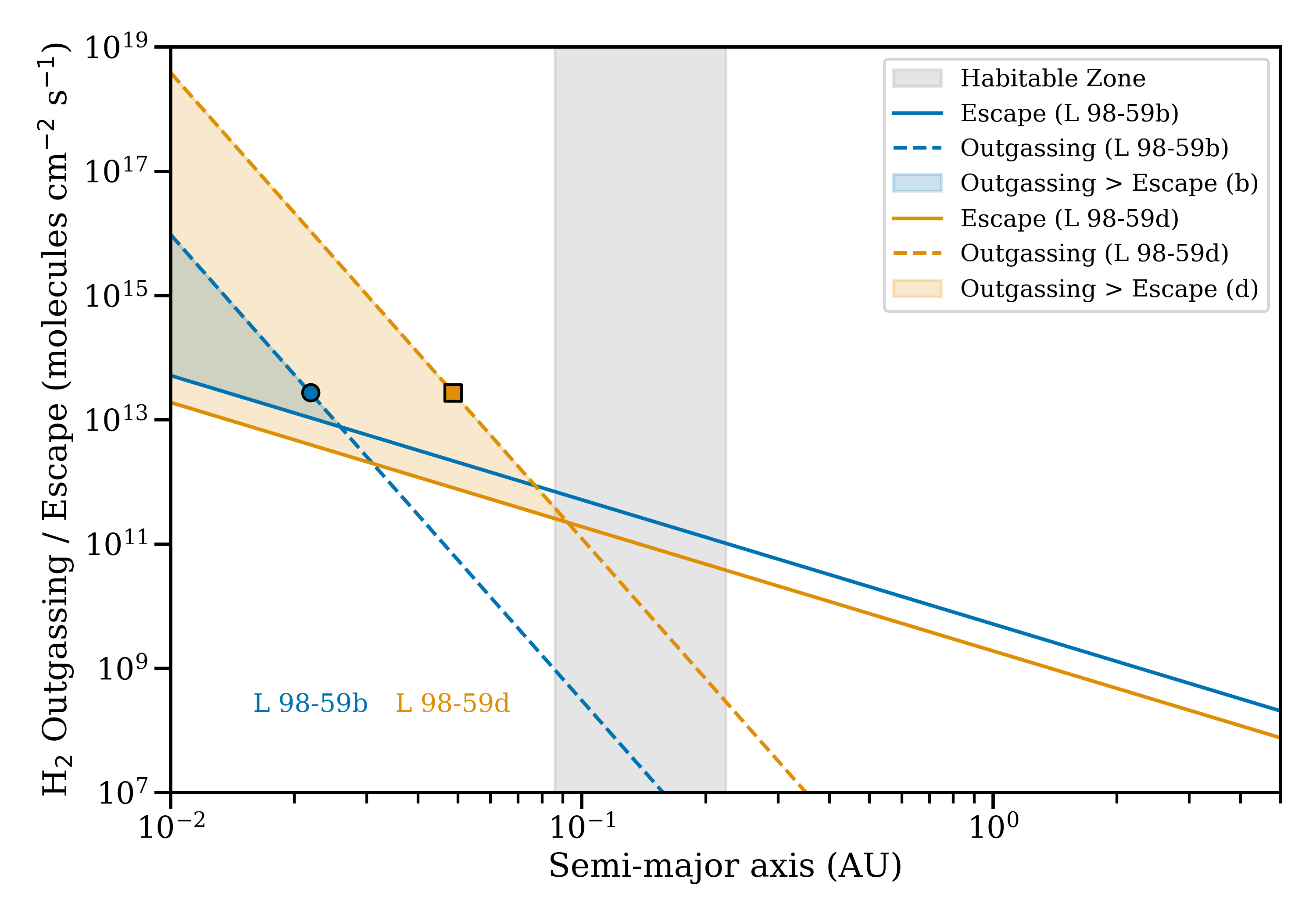}
    \caption{Orbital dependence of hydrogen (\ch{H2}) outgassing and atmospheric escape for planets L 98-59b (blue) and L 98-59d (orange). Solid lines represent energy-limited escape rates, while dashed lines show estimated \ch{H2} outgassing rates driven by tidal heating. Shaded regions highlight the "outgassing zones"—orbital ranges where volcanic \ch{H2} outgassing exceed atmospheric escape, enabling the potential buildup of a secondary \ch{H2}-dominated atmosphere. The grey vertical band marks the conservative habitable zone based on the \cite{Kopparapu2013} criteria. Points indicate the present-day semi-major axes of L 98-59b (circle) and d (square), showing that both planets reside inside their respective outgassing zones. The inner limit of the outgassing zone is not shown in the figure as it lies inner to 0.01 AU.}
    \label{fig:orbital}
\end{figure}

\begin{figure*}
    \centering
    \includegraphics[scale =0.35]{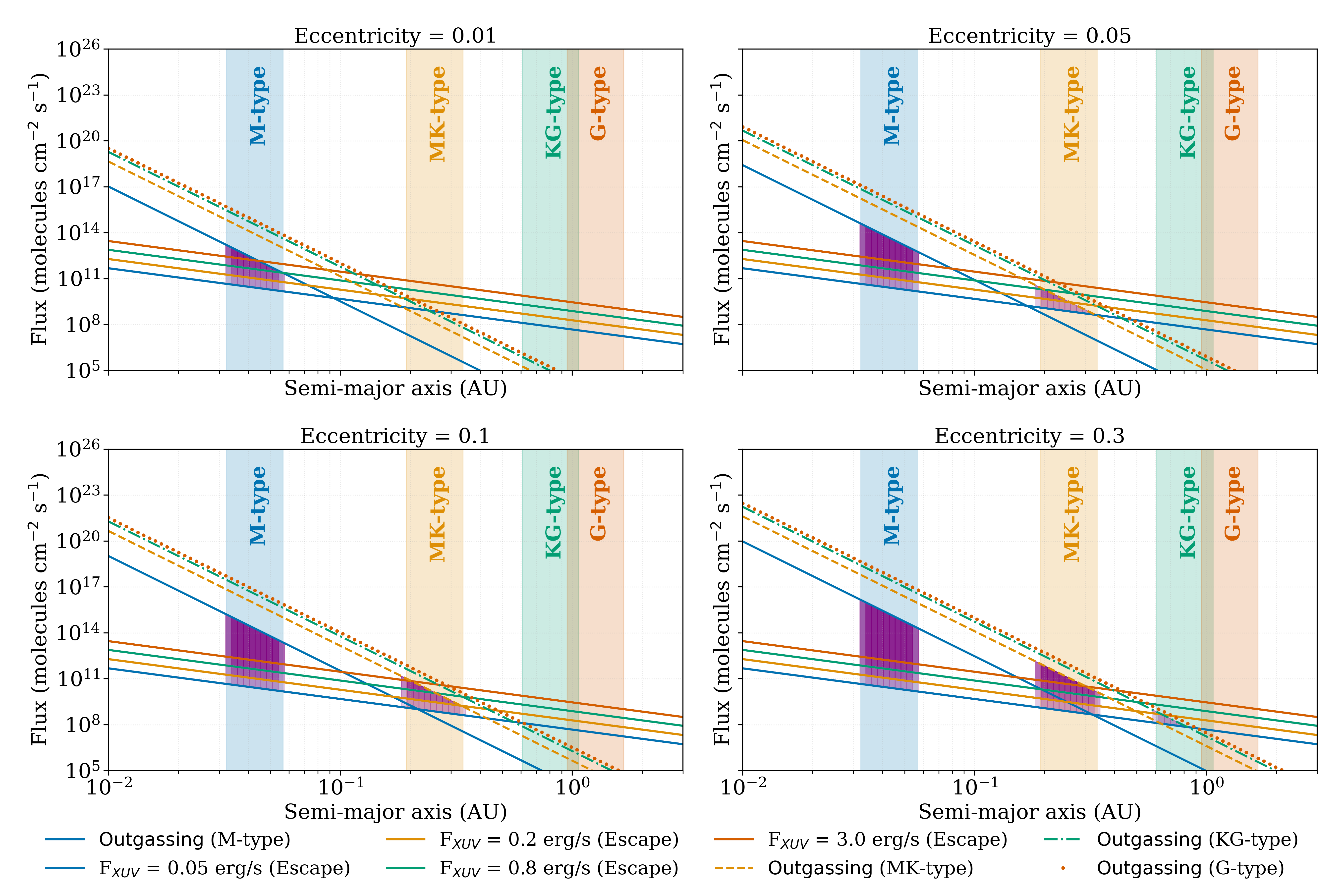}
    \caption{Comparison of outgassing and atmospheric escape fluxes across a range of semi-major axes and eccentricities for different stellar types. Each panel corresponds to a specific orbital eccentricity. Colored bands represent the habitable zones (HZ) for M, K, and G-type stars \citep{Kopparapu2013}. Solid and dashed lines indicate escape and outgassing fluxes, respectively, for different stellar masses. The shaded purple regions highlight where outgassing exceeds escape within the respective HZs, indicating favorable conditions for atmospheric retention.}
    \label{fig:outzone}
\end{figure*}

We consider the area of orbital parameter space where outgassed H$_2$ atmospheres are more likely to persist. We identify an \textit{outgassing zone} (OZ), a region in orbital space where the magmatic outgassing flux of \ch{H2} can exceed escape to space (Figure \ref{fig:outzone}) for plausible $\eta_{\mathrm{og}}\eta_{\mathrm{Heat}}$ and Earth-like outgassing speciation. Interestingly, the OZ lies closer to the host star, where escape rates are expected to be higher. This counterintuitive result arises from the contrasting dependencies of tidal heating and atmospheric escape on orbital distance. Specifically, tidal heating scales as $\dot{E}_{\mathrm{heat}} \propto a^{-7.5}$, while the hydrogen escape flux follows a shallower dependence, $\phi_{\mathrm{esc}} \propto a^{-2}$. In Figure~\ref{fig:orbital}, we illustrate these dependencies for the L 98-59 system. Both L 98-59b and L 98-59d are found within this dominant outgassing zone, suggesting that—provided other conditions discussed in Section \ref{sec:cond} are met—these planets may be capable of sustaining \ch{H2}-rich atmospheres. Our model further predicts that present-day outgassing on TRAPPIST-1b, TRAPPIST-1c, all L 98-59 planets, and LP 791-18d can exceed escape under Earth-like outgassing speciation.
The  tidal heating typically increases faster than escape at smaller \(a\) and hence an inner boundary cannot be defined in the same was as the outer boundary. The inner edge of the OZ is set by the distance at which escape would remove a 16\% by mass \(\ch{H2O}\) inventory in \(<1~\mathrm{Gyr}\) assuming an Earth-sized planet with mantle mass fraction of 0.67 and all water stored in the mantle. For L 98-59, this corresponds to an escape flux of \(2\times10^{14}\) molecules cm\(^{-2}\) s\(^{-1}\), which occurs at $<0.01$ AU. 

The OZ typically lies closer to the star than the habitable zone (HZ; \citealt{Kasting1993}). This is relevant because the formation of an \ch{H2}-rich atmosphere depends not only on \ch{H2} outgassing and escape, but also on the cycling of other chemical species between a planet’s interior and atmosphere, which is affected by the presence of a liquid water ocean which can only exist in the HZ (Section \ref{sec:cond}). Figure \ref{fig:outzone} illustrates how the extent of the outgassing zone varies with stellar type (M, K, and G), F$_\mathrm{XUV}$ levels and eccentricity. We calculated HZ following \cite{Kopparapu2013}, using Equation \ref{eq:hzd}.
\begin{equation}
    d = (\frac{L/L_{\odot}}{S_{\mathrm{eff}}})^{0.5} \mathrm{AU}
    \label{eq:hzd}
\end{equation}
where L is the luminosity of the star, L$_{\odot}$ is the luminosity of the Sun, and S$_{\mathrm{eff}}$ is effective solar flux. For the inner edge of the HZ, we adopt S$_{\mathrm{eff}}$ = 1.06 , which corresponds to the runaway greenhouse limit—where an Earth-like planet would experience complete ocean evaporation. For the outer edge, we use S$_{\mathrm{eff}}$ = 0.325 , corresponding to the maximum greenhouse limit \citep{Kopparapu2013}. These choices aligns with our goal of identifying an overlap between the OZ and the HZ, where surface water can actively regulate atmospheric carbon and oxygen levels, as discussed in Section~\ref{sec:species}. The stellar luminosities are obtained using mass-luminosity relation described in Equation \ref{eq:mass-l} \citep{Duric2004}.
\begin{equation}
\frac{L}{L_\odot} =
\begin{cases}
0.23 \left( \frac{M}{M_\odot} \right)^{2.3}, & M < 0.43\,M_\odot \\
\left( \frac{M}{M_\odot} \right)^4, & 0.43\,M_\odot \leq M < 2\,M_\odot \\
1.4 \left( \frac{M}{M_\odot} \right)^{3.5}, & M \geq 2\,M_\odot
\end{cases}
\label{eq:mass-l}
\end{equation}
where L is the luminosity of the star, L$_{\odot}$ is the luminosity of Sun, M is the stellar mass and M$_{\odot}$ is solar mass. Overall, for an Earth like planet, low-mass M-type stars are the most promising hosts for \ch{H2}-rich atmospheres in the HZ. Higher-mass stars can also support such planets, but typically at higher eccentricities (Figure \ref{fig:outzone}).

\subsection{\ch{H2}-dominated Atmospheres are Accessible to JWST}
\label{sec:snr}
\begin{figure}
    \centering
    \includegraphics[scale = 0.18]{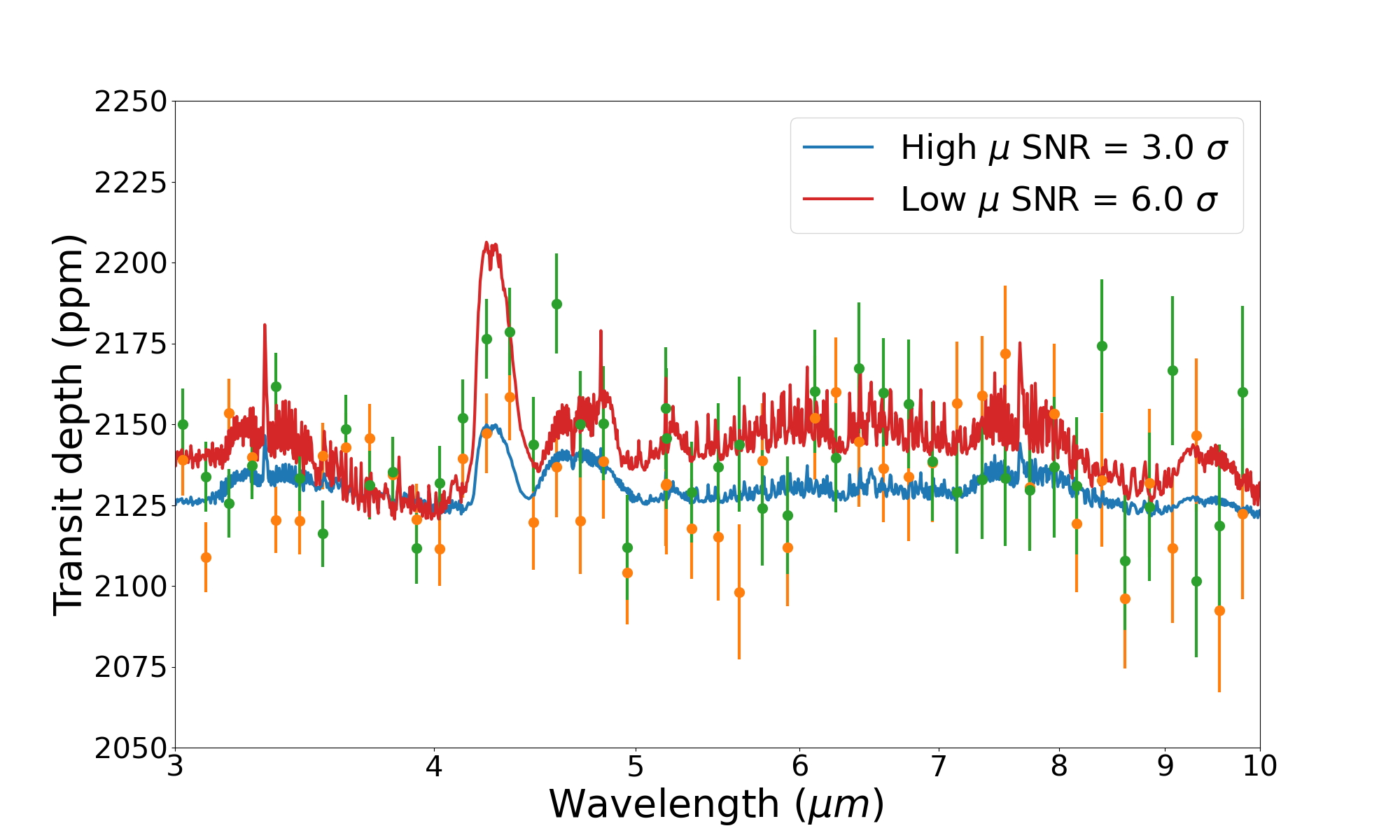}
    \caption{The Figure shows a comparison of spectral simulation of L 98-59 d with two bulk compositions, 100\% \ch{N2} (blue) and 30 \% \ch{N2} + 70 \% \ch{H2} (red). The JWST simulated observations including noise (3 transits, R (wavelength resolution) = 20) for the two respective cases are show in orange (30 \% \ch{N2} + 70 \% \ch{H2}) and green (100\% \ch{N2}). The low and high $\mu$ atmospheres are detected at 6 $\sigma$ and 3 $\sigma$ confidence using 3 transits.}
    \label{fig:spectrum_comp}
\end{figure}

The major advantage of a thin \ch{H2}-dominated atmosphere as a diagnostic for magmatic outgassing is its accessibility to transmission spectroscopy observations, e.g. with JWST. The high concentration of \ch{H2} leads to a larger scale height atmosphere compared to a \ch{N2} dominated atmosphere \citep{miller2009,Burrows_2014}. This larger scale height enhances the strength of absorption features leading to higher transit depths which are more readily observable by JWST. To illustrate this, we compare the detectability of a 10 amu (70\% $H_2$/30\% N$_2$, motivated by \citealt{Banerjee_2024}) and a 28 amu (100\% N$_2$) atmosphere of L 98-59d. We use \texttt{CLIMA} to compute a adiabat-to-isotherm temperature-pressure profile following  \citet{Wogan_2023}. This temperature profile serves as input for the photochemical simulations conducted using the MIT Exoplanet Atmospheric Chemistry (MEAC) model \citep{hu1,Hu2}, which solves for the vertical distribution of atmospheric species under stellar irradiation. We  maintain abiotic \ch{CO2} dominated Earth-like boundary conditions as described in \citet{Hu2}. We then processed this simulated atmosphere through the radiative transfer code petitRADTRANS \citep{Molli_re_2019,blain2024}. We included opacities from major species like \ch{H2O}, \ch{CO2},\ch{CO}, \ch{CH4}, \ch{SO2}, \ch{H2S} and CIA as well as Rayleigh scattering from \ch{H2}, He and \ch{N2}. Figure \ref{fig:spectrum_comp} shows the comparison of transmission depth signals fo the two cases of an atmosphere. 

We predict that the amplitude of transmission signals, relative to the no-atmosphere scenario can reach up to 90 ppm in the near-infrared (NIR) and 60 ppm in the mid-infrared (MIR) for low mean molecular weight atmospheres shown in Figure \ref{fig:spectrum_comp}. In contrast, for high mean molecular weight atmospheres, the relative amplitudes decrease to approximately 30 ppm in the NIR and 20 ppm in the MIR (Figure \ref{fig:spectrum_comp}). The transmission signal from low-$\mu$ atmospheres is above the noise floors of NIRSpec ($<$10 ppm; \citealt{Rustamkulov_2022, Lustig_Yaeger_2023}) and MIRI ($\sim$25 ppm; \citealt{Bouwman_2023}), whereas the signal from high-$\mu$ atmospheres is weaker and approaches the detection limit of these instruments. This accords with extensive past work finding that low mean molecular mass atmospheres are detectable in transmission with JWST in a wide range of scenarios, whereas high mean molecular mass atmospheres are only detectable in extraordinarily favorable circumstances \citep{Greene_2016,LustigYaeger2019}.
To better evaluate the detectability of these atmosphere simulate JWST observations for 3 transits and at a spectral resolution of R = 100 using \texttt{PandExo} \citep{bathala2017}.  The low-$\mu$ spectrum exhibits clear absorption features particularly near $4.3$ and $4.5\,\mu$m corresponding to \ch{CO2}, well above the observational uncertainties, while the high-$\mu$ spectrum shows smaller signals that are closer to the noise level. Further we calculated the SNR of these features following \citet{seligman2023}:
\begin{equation}
\label{eq:SNR}
\langle\mathrm{SNR}\rangle=\sqrt{\sum_{i=1}^{N_\lambda}\left(\frac{m_{1, i}-m_{2, i}}{\sigma_i}\right)^2}
\end{equation}
where m$_1$ is the transmission spectra, m$_2$ is a featureless transmission spectra, and $\sigma$ is the error on the observations. The subscript i is each wavelength point and N$_{\lambda}$ is the total number of resolution elements.
The synthetic spectra reveal that a low mean molecular weight ($\mu$) atmosphere produces prominent spectral features and would prefer an atmosphere over no atmosphere case with a significance of $6\,\sigma$, whereas a high-$\mu$ atmosphere yields more muted features with preference of $3\,\sigma$. 
We also observe \ch{CO2} features to have highest transit depths in Figure \ref{fig:spectrum_comp} with no features of sulfur species which is due to consideration of Earth-like outgassing and surface deposition \citep{Hu2, Ranjan2023c}. These results suggest that JWST will be more sensitive to extended, low-$\mu$ atmospheres.

\subsection{Conditions Required for a Volcanogenic \ch{H2}-dominated Atmosphere}\label{sec:cond}
The major weakness of a thin \ch{H2}-dominated atmosphere as a signpost of volcanic outgassing is that such atmospheres can only be sustained over a narrow corner of geochemical parameter space. Specifically, to sustain a geologically long-lived \ch{H2}-rich atmosphere, the mantle composition and surface conditions must be conducive to \ch{H2} outgassing, the planetary volatile inventory must be adequately large, and heavier gases should not build up in the atmosphere. These conditions mean that we cannot expect all tidally heated exoplanets to host a thin \ch{H2}-rich atmospheres; in other words, the \ch{H2}-rich atmospheres diagnostic for planetary volcanism has a high ``false negative" rate.\footnote{By ``false negative” we mean cases where volcanism is present without a thin, \ch{H2}-rich atmosphere. This can occur if (i) the mantle is relatively oxidized and outgassing favors \ch{H2O}/\ch{CO2} over \ch{H2}; (ii) escape or surface/interior sinks outpace supply; (iii) the hydrogen inventory in the mantle is too small to outgass sufficient \ch{H2} (i.e., $\mu$ is effectively high); or (iv) observational limitations (e.g., clouds/hazes, stellar activity, instrument noise)}
On the other hand, these conditions also mean that detection of a thin \ch{H2}-dominated atmosphere strongly constrains the surface and interior properties of the underlying exoplanet. 

\subsubsection{Volatile Inventory}\label{sec:volinv}

The first and most basic requirement for a geologically long-lived thin H$_2$ atmosphere is adequate H$_2$O volatile inventory. The H$_2$ in an H$_2$-dominated atmosphere is continuously lost to hydrodynamic escape and must be replenished by outgassed H$_2$, ultimately sourced from H$_2$O stored in the interior of the planet. In order to sustain an \ch{H2}-dominated atmosphere, a planet must possess a sufficiently large H$_2$O inventory to maintain an outgassing flux matching or surpassing atmospheric escape rates over geologic time ($\geq 1$ Gyr: \citealt{owen2017}).

We quantify the volatile inventory lifetime, which we define as the time required to fully outgas the available H stored in the interior as a function of outgassing rate and initial volatile inventory:

\begin{equation}
    t_{\mathrm{inv}} = \frac{ N_a}{\mu_{\mathrm{\ch{H2O}}}} \times m_{\mathrm{\ch{H2O}}}M_{\mathrm{P}} \times \frac{1}{\phi_{\mathrm{H_2},P}}
    \label{eq:invenlife}
\end{equation}

where \( t_{\mathrm{inv}} \) is the lifetime of the volatile inventory against escape, \( m_{\mathrm{\ch{H2O}}} \) is the mass fraction of \ch{H2O} in the planet, \( \mu_{\mathrm{\ch{H2O}}} \) is the molar mass of \ch{H2O} in g/mol, respectively. We calculate \( \phi_{\mathrm{H_2},P} \) using Equations \ref{eq:outgflux} and \ref{eq:tidalheat}, exploring \( \eta_{\mathrm{og}}\eta_{\mathrm{Heat}} = 10^{-4} - 1 \). We assumed all hydrogen to be lost as \ch{H2} for simplicity. We consider two limiting endmembers for \( m_{\mathrm{\ch{H2O}}} \). For our low volatile (LV) inventory endmember, we consider an Earth-like water inventory. For our high volatile (HV) inventory endmember, we consider a water-rich basal magma ocean (BMO). 

\paragraph{Earth-like volatile inventory}
 For our LV endmember, we use the 2 $\sigma$ upper bound of the Earth's estimated water capacity (4.41 OM; \citealt{dong2021}). This corresponds to a water mass fraction of \( m_{\mathrm{\ch{H2O}}} = 10^{-3} \). 
 
\begin{figure}
    \centering
    \includegraphics[scale = 0.3]{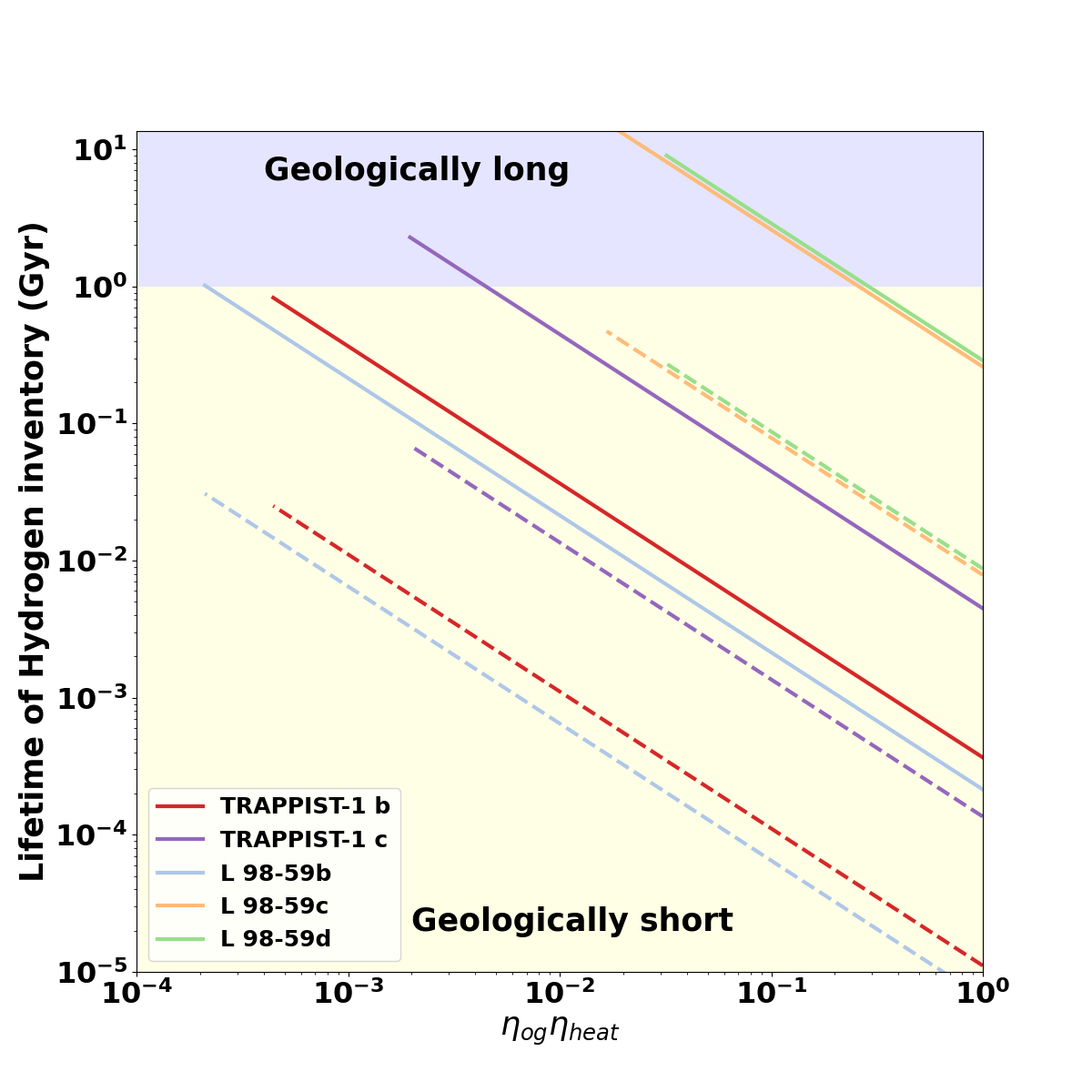}
    \caption{Timescale at which planets having higher outgassing rate then escape rate, will loose all the volatile inventory for 2 extreme cases of mass fraction of \ch{H2O} in the magma (3.3 $\times$ 10$^{-2}$ in solid lines and $1\times10^{-3}$ in dashed lines). The figure presents this result as a function outgassing efficiency varying between 10$^{-4}$ - 1.}
    \label{fig:inventime}
\end{figure}

 We calculate the lifetime of hydrogen inventory on TRAPPIST-1~b/c and L~98-59~b/c/d across varying outgassing efficiencies \( \eta_{\mathrm{og}} \)\( \eta_{\mathrm{heat}} \), using the planetary masses in Table \ref{tab:life}.
Under the LV scenario, none of the considered planets can sustain hydrogen atmospheres for more than 1~Gyr, regardless of $\eta_{og}\eta_{Heat}$. This signifies that even with an optimistic Earth volatile budget, planets with Earth-like volatile inventories are unlikely to sustain \ch{H2}-rich secondary atmospheres over geological time (Figure \ref{fig:inventime}).


Therefore, to sustain an H$_2$-dominated atmosphere over timescales comparable to a planet’s age, a planet must acquire and preserve significantly more $m_{\mathrm{\ch{H2O}}}$ than Earth. In addition to providing the volatiles to sustain outgassing, a higher $m_{H_{2}O}$ also facilitates the high outgassing required to sustain a thin H$_2$-dominated atmosphere against escape by promoting melt production and degassing. First, the presence of water lowers the solidus temperature of silicate rocks, thereby enhancing melt generation within the mantle and crust \citep{asimow2003,Hirschmann2006}. Second, water reduces the viscosity of silicate melts \citep{Hartung2019}, enabling more rapid bubble growth and facilitating efficient degassing. 

\paragraph{Magma ocean-like volatile inventory} For our HV endmember, we consider a water-rich basal magma ocean. Magma oceans (MOs) can act as substantial volatile reservoirs due to the high solubility of hydrogen and water in silicate melts \citep{Bower2021}. However, the lifetime of MOs is short compared to a planet’s full evolutionary timescale. Once solidification occurs, the resulting rock stores much less water than silicate melts, as observed for Earth~\citep{Hirschmann2006}, and the inventory can be lost to the atmosphere and then to space. Basal magma oceans (i.e., where a substantial portion of the mantle is molten) mitigate this problem by having a solid lid sealing the melt from the atmosphere, throttling escape. Basal magma oceans are proposed to have existed on early Earth and Mars \citep{Ballmer2025}, may exist on Venus and Io today \citep{ORourke2020, Aygun2025}, and are anticipated on exoplanets \citep{Boley2023}. In particular, eccentric exoplanets can have long-lived molten mantles due to intense tidal heating \citep{seligman2023, nichollas2025a, nichollas2025b}. 

We therefore adopt a water-rich basal magma ocean as our HV endmember. \cite{kite2021} propose that planets may incorporate up to 3.4~wt\% \ch{H2O} into their magma-filled mantles via endogenous production of H$_2$ from direct accretion of nebular H$_2$. Based on \cite{kite2021}, we assume an inventory of 2-10 $\times$ \(10^{23}\) kg of water (136-680 OM) for a 5 \(M_{\oplus}\) planet. We use the upper bound of 680 OM, corresponding to \( m_{\mathrm{\ch{H2O}}} = 0.034 \), for our HV endmember. We note tentative observational evidence for exoplanets with $m_{\mathrm{\ch{H2O}}} \geq0.01$ \citep{Coulombe2025}, consistent with this scenario.

For the HV endmember, long-lived H$_2$-dominated atmospheres are possible (Figure~\ref{fig:inventime}). TRAPPIST-1~c and L~98-59~c/d can retain hydrogen atmospheres for $>$1 Gyr at relatively low outgassing efficiencies (\( \eta_{\mathrm{og}}\eta_{\mathrm{Heat}} = 10^{-2} \)). Though TRAPPIST-1~c lifetime is $>$ 1 Gyr when \( \eta_{\mathrm{og}}\eta_{\mathrm{Heat}} \lesssim 10^{-2} \), it is still far less than the system’s age \citep{kostov2019} and hence we do not expect TRAPPIST-1c to be currently outgassing.  Thus, L~98-59~c/d appear to be more favorable candidates for sustaining hydrogen-rich secondary atmospheres over geological timescales.


\subsubsection{Speciation of outgassing gas} \label{sec:species}
The second main requirement for the existence of a thin H$_2$-dominated atmosphere is efficient removal or low outgassing of high-$\mu$ species. On Earth, high-$\mu$ species like H$_2$O and CO$_2$ are outgassed at rates much higher than H$_2$ \citep{2017aeil.book.....C}, and our calculations show this to be true across most of parameter space. This means that for the atmosphere to be H$_2$-dominated, either sink mechanisms must be active which remove these high-$\mu$ species, or planetary geochemical conditions must suppress their outgassing. We propose that temperate habitable planets with liquid water oceans are likely to meet this condition, but non-habitable planets require specific conditions (most significantly, outgassing from reduced mantles, e.g \textit{f}O$_2\lesssim$ FMQ-4.4 for L~98-59~d) to achieve this condition\footnote{For context, the mantle of Mercury is estimated to be at \textit{f}O$_2\approx$ FMQ -9  \citep{zolotov2013}}.

\paragraph{Removal of heavier gases on HZ planets} Habitable planets with liquid water oceans can maintain H$_2$-dominated atmospheres even if outgassing is not H$_2$-dominated because of geochemical cycles which scrub high-$\mu$ species like CO$_2$ and H$_2$O from the atmosphere. On a planet with an ocean, the carbonate-silicate cycle regulates atmospheric CO$_2$ \citep{Kasting1993, Hakim2023}, even on tidally-heated planets with high heat fluxes and outgassing rates \citep{Reinhold2025}. Further, plate tectonics may additionally net subduct CO$_2$ from the surface as carbonate \citep{keleman2015}. Even in the absence of silicate weathering, oceans can limit atmospheric CO$_2$ via formation of CO$_2$ clathrate \citep{Levi2017, Ramirez2018}. Similarly, on a temperate planet with T$_{surf}\leq320$K, the upper atmosphere is cool, limiting through condensation and rainout to the ocean the atmospheric H$_2$O inventory \citep{Kasting1988, Wordsworth2013b}. Sulfur-bearing gases like SO$_2$ and H$_2$S are soluble and reactive and scrubbed from the atmosphere by deposition or conversion to aerosols \citep{hu2013, Ranjan2023c}. Consequently, habitable zone planets with liquid-water oceans can plausibly sustain H$_2$-dominated atmospheres even in the face of abundant outgassing of high-$\mu$ species like CO$_2$, H$_2$O, and SO$_2$. While abundant H$_2$ does move the habitable zone outward from the canonical CO$_2$-N$_2$-H$_2$O HZ, the inner edge moves outward by only $\leq4\%$ for H$_2$ mixing ratios up to 50\% v/v, because the greenhouse effect of H$_2$-H$_2$ CIA is masked by H$_2$O absorption \citep{Ramirez_2017}. We therefore propose that planets at the intersection of the outgassing-zone and habitable zone are particularly compelling targets to search for thin H$_2$-dominated atmospheres. 

\paragraph{Removal of heavier gases on non-HZ planets} The situation is more challenging for planets interior to the habitable zone, which are too hot to host liquid-water oceans and the geochemical cycles which regulate CO$_2$, H$_2$O, and SO$_2$/H$_2$S on Earth. On these planets, it is generally assumed that outgassed species remain in the atmosphere unless modulated by photochemistry, surface interactions, or atmospheric escape \citep{Liggins2023, Gkouvelis2025}.
We assessed whether hydrodynamic escape of H$_2$ on outgassing-zone planets interior to the habitable zone is sufficiently vigorous to entrain and remove heavier species such as atomic C, S, and O. To evaluate this possibility, we calculated the crossover mass for our candidate outgassing-zone planets \citep{2017aeil.book.....C}:
\begin{equation}
    m_{\mathrm{c}} = m_1 + \frac{kTF_1}{bgX_1}
    \label{eq:crossovermass}
\end{equation}
\noindent where $m_{\mathrm{c}}$ is the mass threshold below which particles begin to be dragged by the hydrodynamic H wind, $k$ is the Boltzmann constant, $T$ is the exobase temperature (adopted as Earth's $T = 1000$ K; \citealt{Emmert2021}), $F_1$ is the H escape flux, \textit{b} is the binary diffusion coefficient ($b$ for C, S and O through H background, see Equation 15.29 in \citealt{BANKS1973} and Table 1 in \citealt{Zanhle1986} respectively), $g$ is gravitational acceleration, and $X_1$ is H mol fraction. We find that $m_{\mathrm{c}}<m_C, m_O, m_S$ for outgassing zone planets which can plausibly sustain outgassed H$_2$ atmospheres over geologic time. We therefore conclude that over time, heavy species  accumulate in a thin H$_2$ atmosphere as postulated by \citet{Gkouvelis2025}, and therefore in order to sustain a thin H$_2$-dominated atmosphere on a planet interior to the habitable zone, the outgassing must be so deficient in C, S and O outgassing that the atmosphere remains H$_2$-dominated despite accumulation of C, S and O over geologic time. 

\begin{figure}
    \centering
    \includegraphics[scale =0.18]{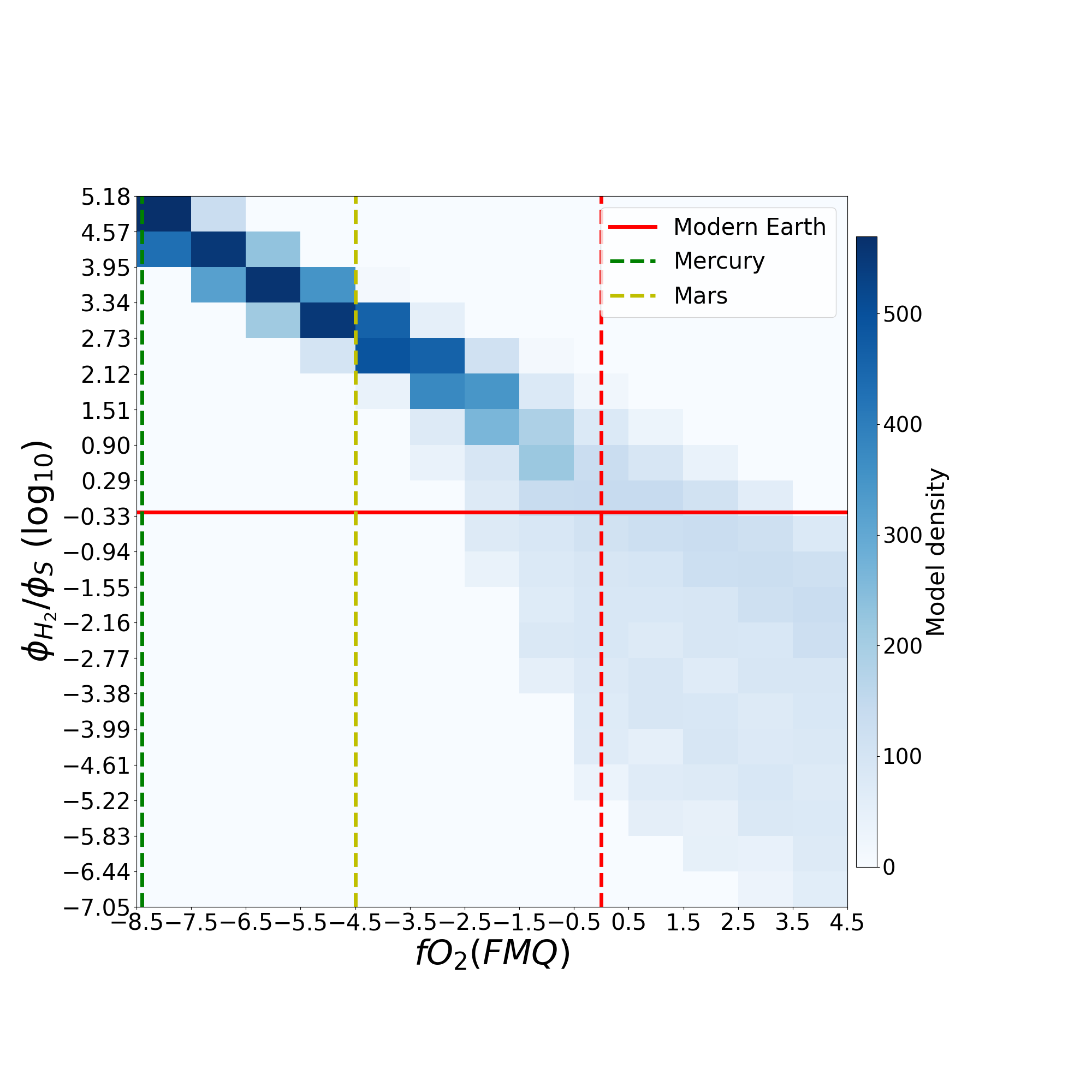}
    \caption{Number of models meeting given $\frac{\ch{H2}}{\ch{S}}$ outgassing flux ratio as a function of fO$_2$ at surface temperature of 1573 K. All other parameters (Table~\ref{tab:parameterraneg}) are marginalized over. Bulk redox states of Mercury\citep{zolotov2013}, Earth \citep{2017aeil.book.....C} and Mars \citep{Doyle_2020} are marked by dashed lines and earth \ch{H2}/S outgassing ratio is shown bu solid line \citep{Hu2023}. }
    \label{fig:H2-S}
\end{figure}

We identify a planet with a reduced mantle as the most likely scenario in which a rocky planet interior to the habitable zone can sustain a thin H$_2$-dominated atmosphere through geologic time. Reduced mantles may occur for planets which had long-lived magma oceans reduced by contact with thick H$_2$ envelopes, as suggested by the radius valley \citep{Luger2015b, KiteFegley2020}. Reduced mantles suppress outgassing of C because under extremely reducing conditions C precipitates out of the melt as graphite instead of releasing to the atmosphere (Section \ref{sec:graph}: \citealt{dasgupta_2013,ortenzi2020,thompson2022}). Reduced mantles suppress outgassing of S (Figure \ref{fig:H2-S}) and N \citep{liggins2,Libourel2003} because under reducing conditions they are more soluble in the melt. Reduced mantles suppress outgassing of O (as H$_2$O) by shifting Equation~\ref{eq:1} in favor of H$_2$. Additionally, O can react with interior reductants \citep{wordsworth_2018}. 

\begin{figure*}
    \centering
    \includegraphics[scale =0.2]{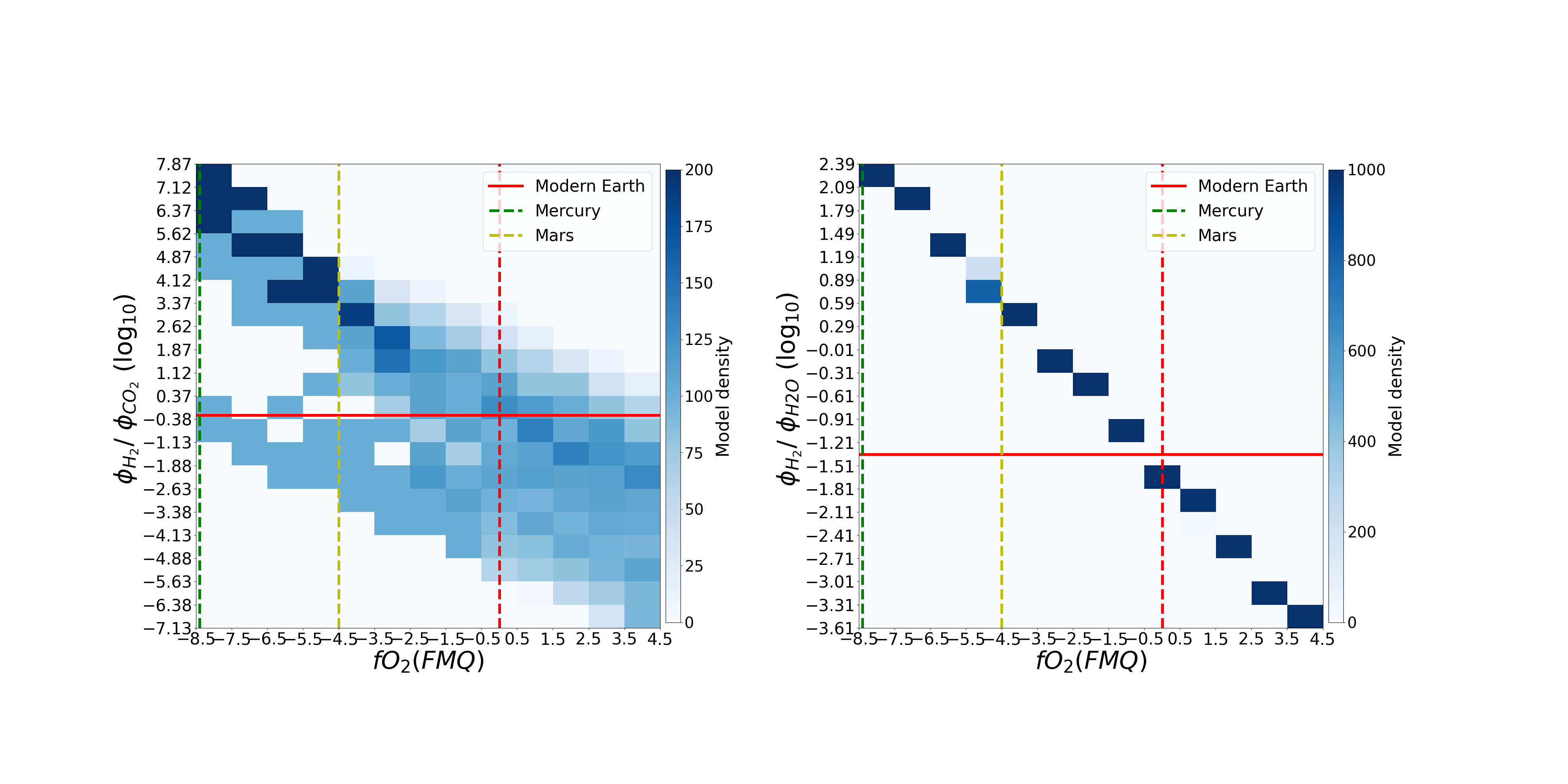}
    \caption{Number of models meeting given $\frac{\ch{H2}}{\ch{CO2}}$ outgassing flux ratio as a function of fO$_2$ at surface temperature of 1573 K. All other parameters (Table~\ref{tab:parameterraneg}) are marginalized over. Red horizontal line shows Modern Earth $\frac{\ch{H2}}{\ch{CO2}}$ outgassing flux ratio for context. Bulk redox states of Mercury\citep{zolotov2013}, Earth \citep{2017aeil.book.....C} and Mars \citep{Doyle_2020} are marked by dashed lines. Solid red lines present Earth outgassing ratios \citep{Hu2023}.  }
    \label{fig:H2co2}
\end{figure*}

We find that for sufficiently reduced mantles (the \textit{f}\ch{O2} threshold is planet-specific), C and S outgassing are sufficiently suppressed and \ch{H2} speciation is higher than \ch{H2O} that the atmosphere remains H$_2$-dominated over geologic time, and enough FeO is available to react away O liberated by H$_2$O photolysis. For a given planet, for each combination of geochemical parameters (Table~\ref{tab:parameterraneg}), there is a steady-state atmospheric pressure for which H$_2$ outgassing matches H$_2$ escape:
\begin{align}
\label{eq:pstedy}
    \phi_{H_{2}, P}(P_{steady})=\phi_{esc}
\end{align}
\noindent Then, under the conservative assumption that H$_2$ but not C and S are lost to escape, we demand:
\begin{align*}
    \frac{\phi_C * 1 \text{Gyr}}{P_{H_{2}, steady}/(g\mu)}\leq0.1\\
    \frac{\phi_S * 1 \text{Gyr}}{P_{H_{2}, steady}/(g\mu)}\leq0.1\\
\end{align*}
\noindent where $\phi_C=\phi_{CO_{2}} + \phi_{CO} + \phi_{CH_{4}}$ and $\phi_{S}=\phi_{SO_{2}}+\phi_{S_{2}}+\phi_{H_{2}S}$. This condition is achieved at different fO$_2$ for different planetary parameters and melt composition. For L98-59d using $\eta_{\mathrm{heat}}\eta_{\mathrm{og}} = 0.001$ \citep{nichollas2025a, Hu2023}, $\chi_{\ch{CO2}_{tot}}=2\times10^{-4}$, $\chi_{\ch{H2O}_{tot}}=3\times10^{-2}$, and $\chi_{\ch{S}_{tot}}=2\times10^{-4}$ (corresponding approximately to Earth MORB enriched in water following \citealt{kite2021} and depleted in C and S due to L 98-59d's lower metallicity; \citealt{saal2002,marty2012,gaillard2021, Demangeon_2021}) we find this condition is achieved for  fO$_2\lesssim\text{FMQ}-4.4$ for both C and S with $P_{H_{2}, steady} \in$ [0.1,P$_{\mathrm{100Myr}}$] bar. For lower C and S inventories, the \textit{f}O$_2$ constraint is relaxed, whereas for larger C and S inventories it is strengthened. 
O outgassing (as H$_2$O) is not suppressed as strongly as C and S outgassing at low \textit{f}O$_2$ (Figure \ref{fig:H2co2}), however, the examples of Mars and Venus show us that O liberated by H$_2$O photolysis can be removed by reaction with reductants in the solid planet \citep{wordsworth_2018}. The O removal power of the solid planet (mantle + crust) can be estimated by quantifying the amount of FeO available to absorb O via the reaction $2FeO + O\rightarrow Fe_2O_3$ \citep{wordsworth_2018}. The condition that there be enough FeO to absorb all O released by H$_2$O outgassing at a rate $\phi_{\ch{H2O},P}$ over 1 Gyr can be written as:
\begin{equation}
    \text{1 Gyr}\times \phi_{\ch{H2O},P}\times4\pi R_{P}^{2}\leq \frac{M_P m_{\mathrm{mantle}} m_{\ch{FeO}}N_A}{\mu_{\ch{FeO}}}
    \label{eq:tfeo}
\end{equation}
\noindent where $m_{\mathrm{mantle}}$ is the mass fraction of mantle, $m_{\ch{FeO}}$ is the mass fraction of \ch{FeO} in the planet, and $\mu_{\ch{FeO}}$ is the molar mass of \ch{FeO}. At \textit{f}O$_2\lesssim$ FQM-4.4, mantle Fe is present overwhelmingly as FeO, not Fe$_2$O$_3$; we therefore take $m_{\ch{FeO}}=10\%$, corresponding to a typical value for solar system mantles \citep{wordsworth_2018}. We take $m_{\mathrm{mantle}}=0.67$ (Earth-like, \citealt{peter1975}). Then, we find that to avoid O$_2$ buildup for an Earth like planet for $\geq$ 1 Gyr, $\phi_{\ch{H2O}}\leq1.8 \times 10{^{13}}$ molecules cm$^{-2}$ s$^{-1}$. This condition is satisfied for \textit{f}O$_2<$FQM-4.4.


The preceding analysis set upper limits on \textit{f}O$_2$ to sustain an H$_2$-dominated atmosphere on a planet interior to the habitable zone. However, to sustain a specifically thin H$_2$-dominated atmosphere, \textit{f}O$_2$ also cannot be too low, because if \textit{f}O$_2$ is too low, then H$_2$ outgassing is so strongly favored over H$_2$O outgassing that P$_{steady}\gtrsim100$ bars. In such cases, it becomes difficult to distinguish between a primary and a secondary atmosphere (Section~\ref{sec:falsepos}). For example, for the L~98-59~d calculation above, the steady-state surface pressure exceeds 17 bar (Figure \ref{fig:press_redx}) for \(\Delta\mathrm{FMQ}\lesssim-4.5\), which would require \(\gtrsim 100\) Myr to escape (Table~\ref{tab:life}), making it difficult to distinguish a secondary outgassed atmosphere from a remnant primary one. In this particular case, the fO$_2$ window for a thin H$_2$-dominated atmosphere is extremely narrow, meaning that detection of one strongly constrains fO$_2$. For other planets and melt compositions, broader thin H$_2$-dominated atmosphere windows are possible; further work is required to explore this parameter space (Section~\ref{sec:caveats}). Here, we simply conclude that thin H$_2$-dominated atmospheres are possible, though potentially over a very narrow parameter space specific to each planetary scenario. 

\begin{figure*}
    \centering
    \includegraphics[scale =0.2]{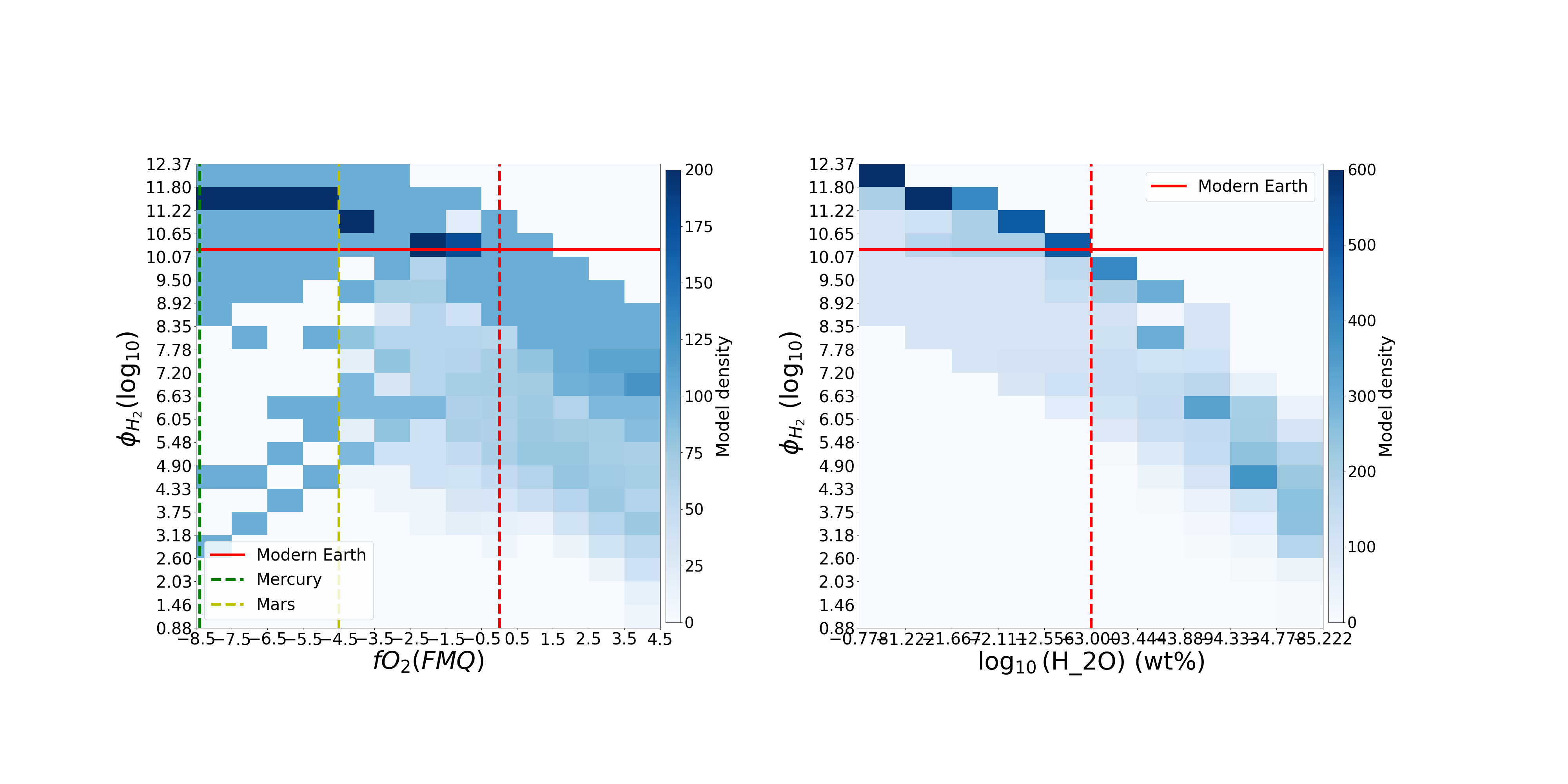}
    \caption{Number of models for the corresponding \ch{H2} outgassing flux as a function of fO$_2$ (left) and \ch{H2O} concentration (right) at surface temperature of 1573 K. Red and Green horizontal lines shows comparison of different mantle conditions with Modern Earth and Mars \ch{H2} outgassing flux. Bulk redox states of Mercury\citep{zolotov2013}, Earth \citep{2017aeil.book.....C} and Mars \citep{Doyle_2020} are marked by dashed lines. Solid red lines amrk earth liek outgassing \citep{Hu2023}. }
    \label{fig:H2-flux1573}
\end{figure*}

\subsection{False Positives for Thin \ch{H2}-Dominated Atmospheres as a Sign of Volcanism} \label{sec:falsepos}
The primary false positive for an \ch{H2}-dominated atmosphere as a sign of volcanic outgassing is primordial \ch{H2}. This false positive is stronger for more massive planets. Mitigating this false positive requires confirming that the atmosphere is specifically a \textit{thin} \ch{H2}-dominated atmosphere, such that its lifetime is geologically short ($<<1$ Gyr) without active replenishment. 

\subsubsection{Accretion and Retention of Primordial \ch{H2}}
Scenarios in which terrestrial planets can retain their natal nebular hydrogen envelopes through geologic time constitute false positives for \ch{H2}-dominated atmospheres as signs of current volcanic outgassing. This scenario is not realized for the Solar System terrestrial planets, but this is because they did not reach their full masses before the disk cleared, meaning they accreted minimal \ch{H2} \citep{Owen2020}. By contrast, the existence of the exoplanet radius gap implies that many if not most sub-Neptune exoplanets finished forming their cores before disk dispersal, accreting substantial nebular envelopes that would constitute a volcanism false positive \citep{Owen2020}. However, the selfsame radius gap implies that these natal envelopes are subject to intense stripping driven by the extreme XUV flux emitted by stars early in their lifetimes, with planets $\lesssim 1.5$ R$_{\oplus}$ completely losing their natal atmospheres and M-dwarf planets being particularly vulnerable to such stripping \citep{Luger2015b, Owen2019, Owen2020}. Therefore, primordial nebular envelopes constitute a possible false positive for \ch{H2}-dominated atmospheres as a marker of volcanic outgassing on an exoplanet, and assessing this false positive requires balancing the atmospheric accretion and escape history to estimate the likelihood of persistence of a primordial envelope for a specific planet.

The evolution and survivability of a primordial \ch{H2}-rich atmosphere depends on three key parameters: the initial gas-to-core mass ratio (GCR; \citealt{evelee2015}) after planet formation, the stellar XUV flux history, and the system age. We estimate the GCR for three endmember disk scenarios (gas-poor dusty, gas-rich dusty and gas-poor dust-free ) following \citet{evelee2015}:
\begin{equation}
\begin{aligned}
\label{eq:dgaspoor}
    \mathrm{GCR}|_{gas-poor, dusty} \simeq 0.06\left(\frac{f}{1.2}\right)\left(\frac{t}{1\, \mathrm{Myr}}\right)^{0.4} \\
    \times\left(\frac{2500\, \mathrm{K}}{T_{\mathrm{rcb}}}\right)^{4.8}\left(\frac{0.02}{Z}\right)^{0.4}\left(\frac{\nabla_{\mathrm{ad}}}{0.17}\right)^{3.4}\left(\frac{\mu_{\mathrm{rcb}}}{2.37}\right)^{3.4}\left(\frac{M_{\text{core}}}{5\, M_{\oplus}}\right)^{1.7}
\end{aligned}    
\end{equation}

\begin{equation}
\begin{aligned}
\label{eq:dgasrich}
    \mathrm{GCR}|_{gas-rich, dusty} \simeq 0.16\left(\frac{f}{3.0}\right)\left(\frac{t}{1\, \mathrm{Myr}}\right)^{0.4} \\
    \times\left(\frac{2500\, \mathrm{K}}{T_{\mathrm{rcb}}}\right)^{4.8}\left(\frac{0.02}{Z}\right)^{0.4}\left(\frac{\nabla_{\mathrm{ad}}}{0.17}\right)^{3.4}\left(\frac{\mu_{\mathrm{rcb}}}{2.37}\right)^{3.4}\left(\frac{M_{\text{core}}}{5\, M_{\oplus}}\right)^{1.7}
\end{aligned}    
\end{equation}

\begin{equation}
\begin{aligned}
\label{eq:gaspoor}
    \mathrm{GCR}|_{gas-poor, dust-free} \simeq 0.16\left(\frac{f}{1.3}\right)\left(\frac{t}{1\, \mathrm{Myr}}\right)^{0.4} \\
    \times\left(\frac{1600\, \mathrm{K}}{T_{\mathrm{rcb}}}\right)^{1.9}\left(\frac{0.02}{Z}\right)^{0.4}\left(\frac{\nabla_{\mathrm{ad}}}{0.17}\right)^{3.3}\left(\frac{\mu_{\mathrm{rcb}}}{2.37}\right)^{3.3}\left(\frac{M_{\text{core}}}{5\, M_{\oplus}}\right)^{1.6}
\end{aligned}        
\end{equation}

\begin{figure*}
    \centering
    \includegraphics[scale = 0.5]{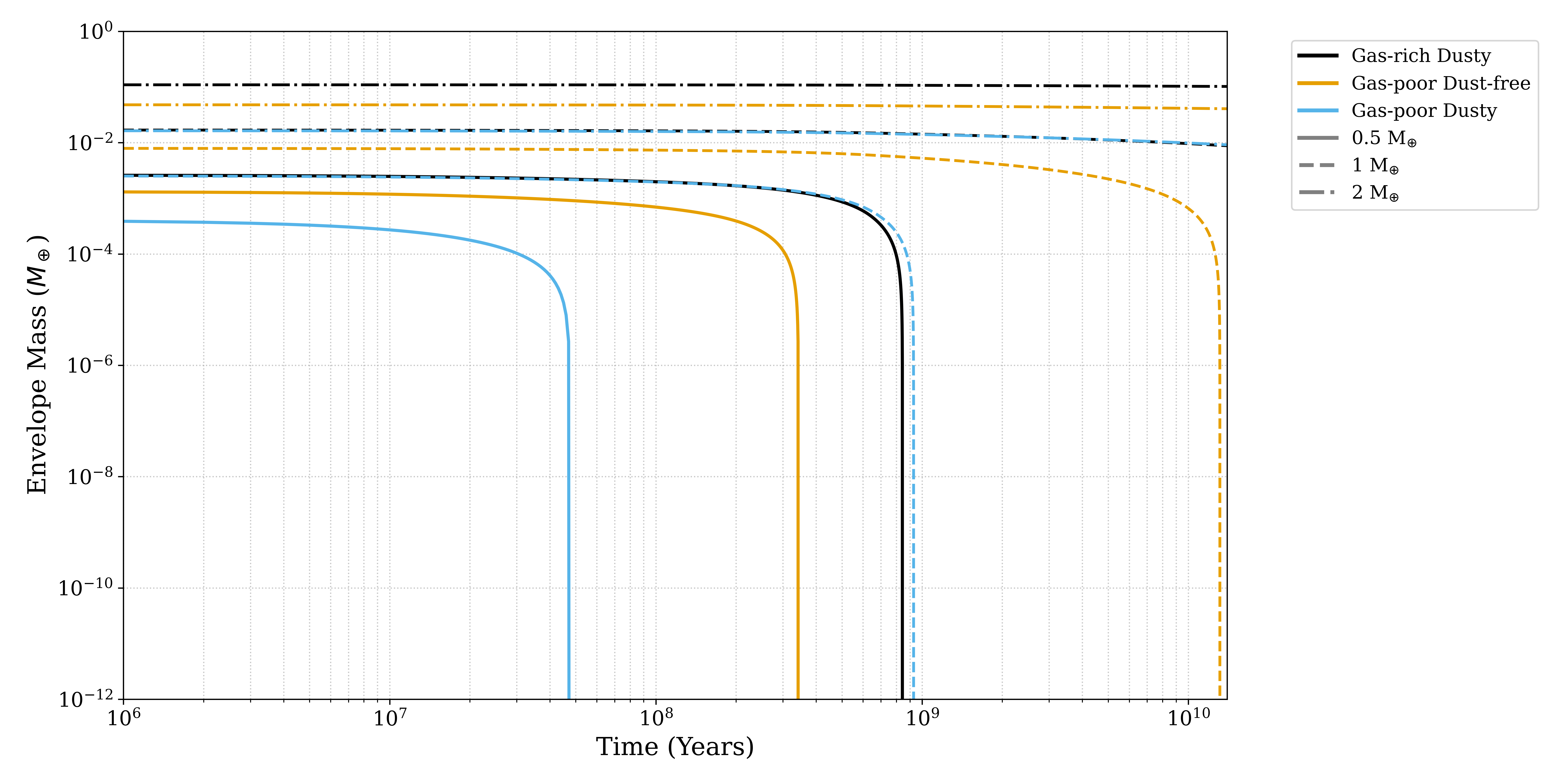}
    \caption{Time evolution of atmospheric envelope mass for hypothetical rocky planets with core masses of 0.5\,M$_\oplus$, 1\,M$_\oplus$, and 2\,M$_\oplus$ (solid, dashed, and dash-dotted lines, respectively), orbiting at 0.1\,AU and radius calculated using mass radius relation of Earth-like rocky planets \citep{Zeng2019}. Curves represent evolution of atmospheric mass under energy-limited atmospheric escape under evolving stellar XUV flux. Initial envelope masses are derived from different disk scenarios: gas-rich dusty (black), gas-poor dusty (blue), and gas-poor dust-free (orange).}
    \label{fig:masslossev}
\end{figure*}

We simulate envelope mass over time for 3 planets with $M_{\rm core}=0.5,1,2 M_{\oplus}$, $a=0.1$ AU and radius from the \citet{Zeng2019} mass-radius relation for Earth-like rocky planets (Figure \ref{fig:masslossev}), including nebular accretion and XUV-driven escape. To calculate nebular accretion, we adopt disk-specific values of f = 1.2 (gas-poor, dusty), 3.0 (gas-rich, dusty) and 1.3 (gas-poor, dust free) \citep{evelee2015}. $T_{\rm rcb} = 2500$~K for the dusty disk models and 1600~K for the gas poor dust-free case and $\nabla_{\rm ad} = 0.17$ \citep{evelee2015}. The value for $\mu_{\rm rcb} = 2.37$ are fixed across all scenarios following \citet{evelee2015}. We assume the solar metallicity of $Z = 0.02$ for these simulations, which sets the initial envelope H$_2$ and He mass fractions $m_{\mathrm{\ch{H2}}}$ and $m_\mathrm{\ch{He}}$ \citep{evelee2015}:
\begin{equation}
\label{eq:h2_mass}
    m_{\mathrm{\ch{H2}}} = \frac{1 - Z}{1.4}=0.7,
\end{equation}
\begin{equation}
\label{eq:He_mass}
    m_{\mathrm{\ch{He}}} = \frac{0.4(1 - Z)}{1.4}=0.28,
\end{equation}
\begin{equation}
\label{eq:mu}
    \mu = \frac{1}{0.5 m_{\mathrm{\ch{H2}}}+ 0.25m_{\mathrm{\ch{He}}}+ 0.6Z}=2.31
\end{equation}

To calculate escape use the time evolution of stellar XUV irradiation using the \texttt{STELLAR} module in the \texttt{VPLanet} suite \citep{Barnes_2020}, incorporating evolutionary tracks from \citet{baraffe2015}. We adopt an XUV saturation fraction of $L_{\rm XUV}/L_{\rm bol} = 0.002$, a saturation timescale of 650 Myr, and stellar mass $M_\star = 0.35\,M_\odot$, similar to L~98-59 \citep{fromont2023}.

Figure~\ref{fig:masslossev} demonstrates the strong dependence of primordial atmosphere survival timescale on planetary mass. Lower-mass planets not only accrete smaller atmospheric envelopes but also lose them more rapidly due to lower gravity and weaker retention. As a result, planetary mass becomes a key discriminator between primordial and secondary atmospheres. For a 0.5\,M$_\oplus$ planet orbiting a 0.35\,M$_\odot$ star at 0.1\,AU, we find that in all three accretion scenarios, the primordial atmosphere is lost within 1\,Gyr, and primordial atmospheres are ruled out on such worlds (Figure \ref{fig:masslossev}). In contrast, for a 2\,M$_\oplus$ planet, all three accretion scenarios result in retention timescales exceeding the age of the universe, and primordial atmospheres cannot be ruled out (Figure \ref{fig:masslossev}). A 1\,M$_\oplus$ planet is transitional between these regime, being able to retain its envelope over 1-10\,Gyr depending on the scenario (Figure \ref{fig:masslossev}). Note that these mass thresholds are not absolute, and depend on the particulars of the planetary system under consideration.\\

 To distinguish the atmospheric origin in such ambiguous cases, we introduce the present-day atmospheric lifetime:
\begin{equation}
    \tau_{\mathrm{life}} = \frac{M_{\mathrm{env}}}{\dot{M}} \, ,
    \label{eq:tau_life}
\end{equation}
where \( \tau_{\mathrm{life}} \) is the lifetime of the observed atmosphere under the current escape rate \( \dot{M} \) (Equation~\ref{eq:els}). If $\tau_{\mathrm{life}}$ $<$ 100 Myrs (geologically short; \citealt{owen2017}), then observing the primordial atmosphere within the narrow transitional loss phase is statistically unlikely and it is instead likely to be a secondary atmosphere, where as for $\tau_{\mathrm{life}}$ $>$ 100 Myrs a primordial atmosphere cannot be ruled out. To distinguish between a primordial and secondary atmosphere for ambiguous cases like a 1\,M$_\oplus$ planet, we must use precise mass-radius constraints to tightly constrain $M_{env}$ (e.g., \citealt{Cadieux2025}), with stellar XUV flux measurements, using the present-day atmospheric lifetime, \( \tau_{\mathrm{life}} \), as a diagnostic. As a reference, we estimate the measurement precision needed to identify a secondary atmosphere under conservative assumptions. We consider a 1\,M$_\oplus$, 1\,R$_\oplus$ planet at 0.01\,AU around  L~98-59 (F$_{\mathrm{XUV}}$ = 5530 erg cm$^{-2}$ s$^{-1}$, \citealt{muscles_extension}) and calculate the escape rate \( \dot{M}  = 1.12 \times 10^{9} \) g/s using the energy-limited escape formalism \citep{Hu2023}. For \( \tau_{\mathrm{life}} = 100\,\mathrm{Myr} \), we obtain an envelope mass fraction of \( f_{\mathrm{env}} =5.9 \times 10^{-4} \).

Using this \( f_{\mathrm{env}} \) value, we apply the scaling relation from \citet{Lopez_2014} to estimate the envelope radius contribution:

\begin{equation}
    \begin{aligned}
    R_{\mathrm{env}} = R_{\mathrm{p}} & - R_{\mathrm{core}} - R_{\mathrm{atm}} = 2.06\,R_{\oplus}\left(\frac{M_{\mathrm{p}}}{M_{\oplus}}\right)^{-0.21} \\
    & \times \left(\frac{f_{\mathrm{env}}}{5\%}\right)^{0.59}\left(\frac{F_{\mathrm{p}}}{F_{\oplus}}\right)^{0.044}\left(\frac{\mathrm{age}}{5\,\mathrm{Gyr}}\right)^{-0.18}
    \end{aligned}
    \label{eq:f_Env}
\end{equation}

We use this methodology to calculate the required precision  of radius to analyse the origin of the atmosphere for a  1\,M$_\oplus$ case. As previously mentioned, we estimate that a lifetime threshold of \( \tau_{\mathrm{life}} = 100\,\mathrm{Myr} \) corresponds to an envelope mass fraction of \( f_{\mathrm{env}} = 5.9 \times 10^{-4} \). 
Using this value in Equation~\ref{eq:f_Env}, we find that the associated atmospheric radius contribution (R$_{\mathrm{env}}$ = 0.029 R$\oplus$) corresponds to a required 3$\sigma$ measurement precision of 2.9\%. 
Therefore confirming a thin secondary atmosphere on planets is in principle possible through high-precision measurements of planetary mass and radius, but higher precision is required than has been achieved for most of the candidate volcanically active worlds.

\section{Discussion}

\subsection{Thin H$_2$-Dominated Atmospheres Are Possible on Tidally-Heated Terrestrial Planets}
\label{subsec:thinH2_tidallyheated}
H$_2$-dominated atmospheres represent the best-case scenario for characterization of terrestrial exoplanet atmospheres via transmission spectroscopy, because their low scale heights amplify spectral features by an order of magnitude relative to high-$\mu$ atmospheres \citep{miller2009,Burrows_2014}. However, past work argued that H$_2$-dominated atmospheres on terrestrial exoplanets are unstable because Earth-like H$_2$ outgassing cannot keep up with atmospheric escape \citep{liggins1, Hu2023}. This would imply that terrestrial planet atmospheres are high-$\mu$, rendering them observationally expensive to characterize \citep{Fortney_2013}. 

Here, we propose that tidally-heated exoplanets in eccentric orbits are capable of enhanced volcanic outgassing which can sustain an H$_2$-dominated atmosphere despite efficient escape, provided 4 main conditions are met:
\begin{enumerate}
    \item $e$ high enough to support high tidal heat fluxes, in excess of Earth's total interior heat flux (Section \ref{sec:tidal})
    \item Presence of a BMO (Section \ref{sec:volinv})
    \item Planetary $m_{H_{2}O}\gtrsim$ 1 \%  (Section \ref{sec:volinv})
    \item Sufficiently reducing melt to suppress S, C and O outgassing or operation of geochemical sinks to prevent accumulation of S, C, and O-bearing species (e.g., hydrological cycle; Section \ref{sec:species})
\end{enumerate}
\noindent Condition (1) is required to permit high outgassing that can compete with escape. Conditions (2) and (3) are required to have enough interior H to sustain the high outgassing rates needed to counterbalance escape. The required $m_{H_{2}O}$ is planet-specific, but in the range of 10$^{-3}$ - 10$^{-2}$ for the planets we have considered (Table~\ref{tab:life}). Condition (4) is required for outgassing of high-$\mu$ species to be suppressed enough for the atmosphere to remain H$_2$-dominated over geologic time. 

These conditions are stringent but not impossible. Basal magma oceans are anticipated as theoretical possibilities for tidally-heated exoplanets \citep{Henning_2009, Driscoll2015, seligman2023, farhat2024, nichollas2025a}. There is observational evidence for planets with water-rich interiors \citep{nichollas2025b, ross2025, Coulombe2025}; this water could be exogenously delivered \citep{Bitsch_2019, Shlecker2021} or  endogenously produced \citep{kite2021} and stored in the interior \citep{Luo2024}. A wide range of mantle \textit{f}O$_2\in[\text{FQM}-9,\text{FQM}+5]$ is observed in the Solar System (Mercury, \citealt{zolotov2013} to Venus, \citealt{Wordsworth2016}), and an even larger range is possible for exoplanets due to different chemical composition of their progenitor disks \citep{wordsworth_2018, Putirka2021, Guimond2023b, Guimond2024,Cioria_2024}. It is therefore \textit{a priori} plausible for such conditions to exist. 

Detection of a thin H$_2$-dominated atmosphere would be exciting because it would comprise evidence of active volcanism on the underlying planet, as escape would otherwise geologically quickly erode the atmosphere (Table \ref{tab:life}). This is important because volcanism is otherwise projected to be extremely difficult to detect on terrestrial exoplanets, detectable only transiently during the largest volcanic events for Earth-analog worlds \citep{Hu2,Misra_2015,Ostberg_2023}. By contrast, an H$_2$-dominated atmosphere can be readily detected with a few transits by JWST, if present (Section~\ref{sec:snr}). The main false positive of primordial H$_2$ for an H$_2$-dominated atmosphere as a signpost of outgassing can be mitigated by mass-radius measurements to confirm whether the planet is specifically a thin atmosphere, with $P\leq P_{100\text{Myr}}$. The existence of such an atmosphere would additionally imply the existence of a water-rich basal magma ocean (Section~\ref{sec:volinv}), constraining the interior structure and evolution of the planet (e.g., \citealt{nichollas2025a}). Lastly, the existence of a thin H$_2$-dominated atmosphere would constrain \textit{f}${O_{2}}$. An upper limit on \textit{f}O$_2$ arises because for a high-enough \textit{f}O$_2$, $\phi_{H_{2},P}<\phi_{esc}$ even for high $m_{H_{2}O}$ and the atmosphere would not be stable. For planets interior to the habitable zone with no surface liquid water, an additional lower limit arises for \textit{f}O$_2$ because if \textit{f}O$_2$ is too low, $P_{\mathrm{Steady}}\!\gg\!P_{100\,\mathrm{Myr}}$, where $P_{100\,\mathrm{Myr}}$ is the surface pressure of atmosphere which would be lost to space in 100 Myr (Section~\ref{sec:volinv}). In summary, detection of a thin H$_2$-dominated atmosphere on a terrestrial exoplanet would strongly constrain its interior state.

Identification of an H$_2$-dominated atmosphere on a terrestrial planet would additionally be important because it would generally enable detailed characterization of the atmosphere and planet below. The trace gas composition of terrestrial exoplanet atmospheres may probe phenomena as diverse as hydrology, impacts, mantle redox, and the presence of life. However, forward modeling studies show that trace gases on terrestrial are  undetectable with JWST for all but (1) highly favorable targets with very small, very close host stars like the TRAPPIST-1 system, and (2) planets with H$_2$-dominated atmospheres and concomitant large scale heights (e.g., \citealt{ Sousa_Silva_2020, Zhan2021, Mikal-evans2022}). Identification of a terrestrial planet with an H$_2$-dominated atmosphere would therefore open the gateway to a broad range of exoplanetary science. 

\subsection{Thick \ch{H2}-Dominated Atmospheres Are Not Necessarily Primordial}

Previous work has highlighted the possibility of primordial H$_2$ atmospheres on terrestrial planets \citep{Owen2020}. Here, we have shown that how that under specific interior and orbital conditions, volcanically sourced hydrogen atmospheres can be sustained over Gyr timescales. This means that detecting \ch{H2}-dominated atmospheres does not necessarily imply the presence of a primordial envelope. Instead, such atmospheres may signal ongoing or past volcanic activity, especially on tidally heated, water-rich planets. Moreover, our framework can guide observational target selection: planets that satisfy the conditions we outline—such as TRAPPIST-1c or L98-59~d—represent promising candidates for detecting volcanic \ch{H2} atmospheres. 

\begin{figure}
\centering
\includegraphics[scale = 0.5]{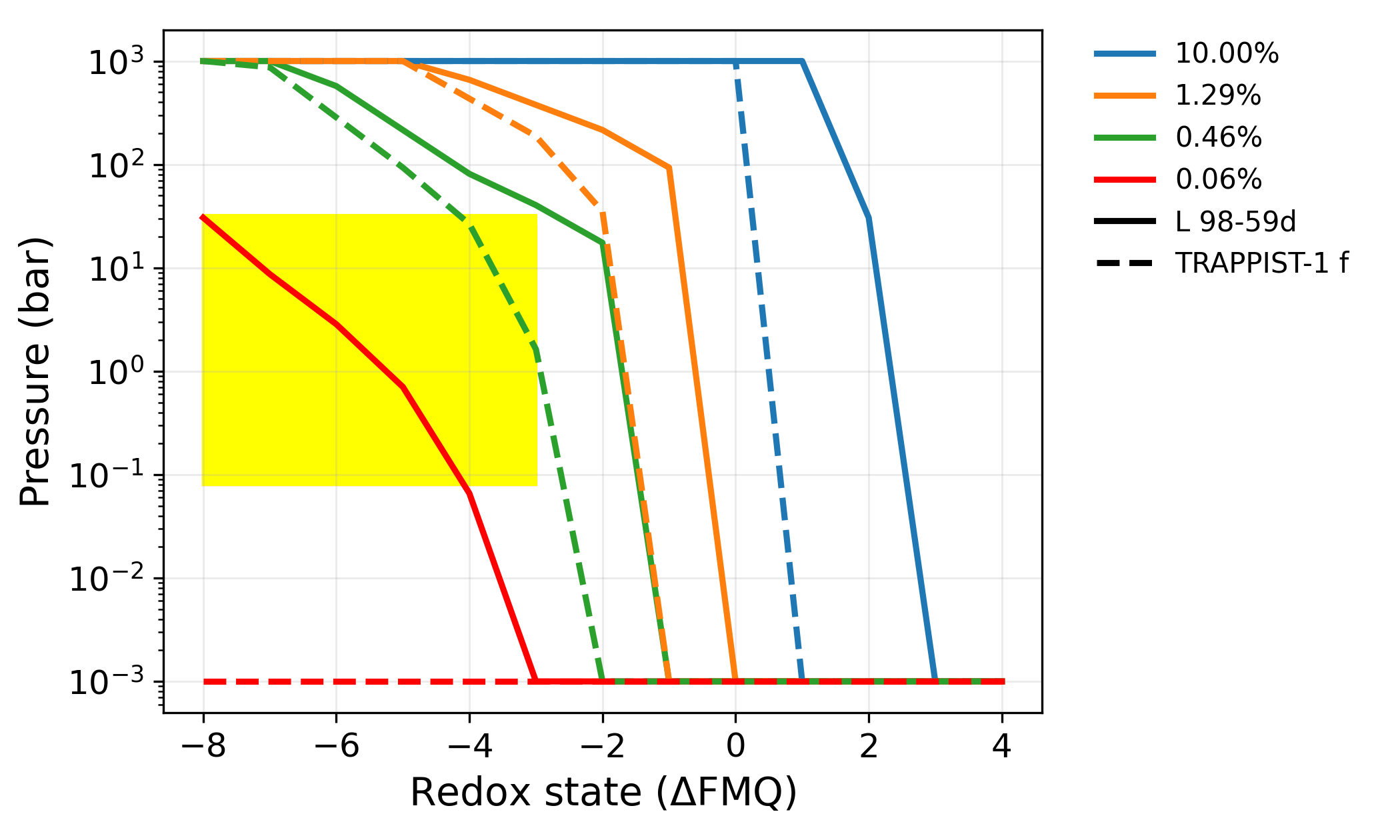}
\caption{Colored curves show the equilibrium surface pressure from the outgassing–escape balance as a function of mantle redox state ($\Delta\mathrm{FMQ}$) for four $m_{\mathrm{\ch{H2O}}}$ (10.00\%, 1.29\%, 0.46\%, 0.06\%). Solid lines correspond to L~98–59,d; dashed lines to TRAPPIST-1,f. The yellow box highlights the observationally relevant regime bounded below by $0.1$~bar (detectability threshold) and above by the surface pressure equivalent to the amount of H$_2$ that would be lost in $100$~Myr at the adopted escape rate. Here we fixed $\chi_{\ch{CO2}}$ = 600 ppm and $\chi_{S}$ = 1000 ppm using MORB volatile inventory \citep{gailard_framework} }
\label{fig:press_redx}
\end{figure}

While we have focused on thin H$_2$-dominated atmospheres as diagnostics of volcanic outgassing, it is possible for volcanic outgassing to generate thick atmospheres as well. For planets like L~98-59~d, with large $m_{H_{2}O}\gtrsim 1\%$ and low \textit{f}O$_2\lesssim\text{FMQ-4.5}$, the outgassing–escape steady state (\(\phi_{\ch{H2},P}=\phi_{\mathrm{esc}}\)) is achieved at very high surface pressures (\(P_{\mathrm{Steady}}\gtrsim 10^{3}\,\mathrm{bar}\)) (Figure \ref{fig:press_redx}). This is because \ch{H2} degassing is regulated by pressure while the \ch{H2}-\ch{H2O} equilibrium is most sensitive to \textit{f}O$_2$; consequently, a large overburden pressure is required to throttle $\phi_{\ch{H2},P}$ to $\phi_{\mathrm{esc}}$ under reducing conditions. Indeed, achieving a specifically thin H$_2$ atmosphere which can be unambiguously attributed to outgassing requires that \textit{f}O$_2$ not be too low, with the specific threshold value dependent on the individual planetary scenario. This suggests that thick H$_2$ atmospheres on tidally-heated planets have an ambigious origin: they may be primordial, or they may be outgassed from a reducing, water-rich mantle.

We consider surface pressures in the range of $0.1\,\mathrm{bar}$ - P$_{\mathrm{100Myr}}$ (yellow box in Fig.~\ref{fig:press_redx}). The $0.1\,\mathrm{bar}$ lower limit is our detectability threshold for a visible atmosphere \citep{dewit2013}, and the upper limit is set by escape-driven $\mathrm{H_2}$ loss over $100\,\mathrm{Myr}$. Moreover, the outgassing escape steady state tends to occur at comparatively oxidizing melt conditions, which favor \ch{H2O}-dominated speciation. Together, these constraints imply that detectable, long-lived atmospheres are most likely when both the bulk \ch{H2O} inventory and the outgassing flux are moderate—not so large that they push pressures above the upper bound or drive the system toward oxidizing, \ch{H2O}-dominated compositions—thereby enabling lifetimes \(\gtrsim 1~\mathrm{Gyr}\). Further work is required to determine observational discriminants between thick, outgassed H$_2$ atmospheres and primordial envelopes.

\subsection{Suggestions for Observations}
\label{subsec:obs_implications}
We propose a three-step procedure to search for thin, \ch{H2}-dominated atmospheres on terrestrial planets. First, use nominal mass, radii, and eccentricity constraints to identify observationally accessible candidate eccentric terrestrial exoplanets; this has already been done \citep{quick2020,seligman2023}, and new candidates are continuously being identified (e.g., \citealt{Damasso2022,brady2025}). Second, use precision mass-radius-eccentricity constraints derived from radial velocity monitoring and photometric transits to confirm that the candidates are actually eccentric (e.g., \citealt{dreizler2024,xue2025,Cadieux2025}) as many nominally eccentric planets are actually consistent with circular orbits within observational error. Additionally, use the precision mass-radius measurements to confirm whether $f_{env}$ is low enough that  $P_{surf}\leq P_{100\text{Myr}}$, ruling out the primordial atmosphere false positive scenario (Section~\ref{sec:falsepos}). Third, measure a low-precision JWST spectrum to check for an H$_2$-dominated atmosphere  via scale-height–driven feature amplitudes and \(\ch{H2\!-\!H2}/\ch{H2\!-\!He}\) CIA. Notably, constraints on $\mu$ are measured as part of standard atmospheric reconnaissance for observationally-favorable small exoplanets (e.g., \citealt{Gao_2023,Damiano_2024,Alderson2024, Glidden2025}), so in many cases this step exploit existing measurements with no new JWST time required. Additionally, note that the required JWST observable ($\mu$) is much more confidently constrained via transmission spectroscopy \citep{miller2009,Benneke_2012} than detection of specific molecular species due to degenerate absorption features \citep{Line_2012,Welbanks_2019}, thereby sidestepping many of the challenges in attempting to interpret transmission spectra (e.g., \citealt{niraula_2025,welbanks_2025}). Our ordering of steps 2 and 3 reflects the assumption that JWST time is more expensive than precision RV/photometric transits, but in practice these steps can be done in either order (Section~\ref{sec:casestudy}).

Counterintuitively, far-orbiting planets are less likely to sustain volcanogenic H$_2$-dominated atmosphere against escape, because tidal heating decreases more quickly with $a$ than escape does. We use this fact to delineate a novel ``outgassing zone" to prioritize targets to search for H$_2$-dominated atmospheres (Figure~\ref{fig:orbital}; Section~\ref{sec:OZ}). Like HZ, OZ limits are not strict, and depend somewhat on assumptions regarding planetary parameters and processes (Section \ref{sec:OZ}). However, like the HZ, the OZ suffices in its purpose of helping prioritize planets for detailed observational investigation. 

We identify planets at the intersection of the HZ and OZ as the most promising candidates for hosting thin H$_2$-dominated atmospheres. Outside the HZ, the absence of surface liquid water removes known efficient geochemical sinks, allowing heavy species (e.g., \ch{CO2}, \ch{SO2}) to accumulate and push atmospheres toward high \(\mu\); this narrows the space where thin \ch{H2} states exist. By contrast, HZ planets that plausibly host surface water can regulate heavy gases via the hydrological and carbonate–silicate cycles and generally experience lower \(\phi_{\mathrm{esc}}\), which raises \(P_{100\,\mathrm{Myr}}\) and makes it easier for \(P_{\mathrm{Steady}}\) to fall within the thin window \([0.1~\mathrm{bar},\,P_{100\,\mathrm{Myr}}]\) even for modest outgassing.

\subsection{Nondetections of Thin \ch{H2}-Dominated Atmospheres on Eccentric Terrestrial Planets Constrain Their Interior State}
\label{subsec:nondetections_eccentric}
Nondetections of thin H$_2$-dominated atmospheres on eccentric terrestrial planets are informative, because they constrain their interior state. Strongly tidally-heated planets with water-rich, reducing mantles \textit{should} host H$_2$-dominated atmospheres (Section~\ref{sec:cond}). Therefore, nondetection of such atmospheres implies the absence of at least one of these conditions, providing a joint limit on $\eta_{og}\eta_{Heat}$, $m_{H_{2}O}$, and \textit{f}O$_2$. Outgassing from water-rich reducing melts are potential false positives for reducing bioindicator gases, including CH$_4$ \citep{Wogan_2020, Liggins2023}, PH$_3$ \citep{Sousa_Silva_2020, Bains_2021}, and NH$_3$ \citep{Huang_2022, Liggins2023}. Interior models can constrain $\eta_{og}\eta_{Heat}$ (e.g., \citealt{nichollas2025a, Gkouvelis2025}), resulting in tighter constraints on $m_{H_{2}O}$ and \textit{f}O$_2$. Nondetection of H$_2$-dominated atmospheres on planets with high tidal heating would indicate that planets with the required melt conditions are rare, thereby disfavoring these false positive scenarios. 


We emphasize that the ability of nondetections of H$_2$-dominated atmospheres to inform their interior state is specific to eccentric planets with high tidal heating (Eq.~\ref{eq:tidalheat}). Nondetections of H$_2$-dominated atmospheres on planets in low-$e$ orbits can be simply explained by low interior heat flux (e.g., \citealt{Hu2023}), so it is important to confirm and precisely measure high eccentricity. This could be obtained combining radial velocities, transit-timing variations \citep{Cadieux2025}, and, when available, photo-eccentric or secondary-eclipse timing, with multi-epoch baselines long enough to average over stellar activity and to sample apsidal precession \citep{wit2012,Csizmadia2019}. Careful treatment of activity-driven RV systematics and TTV degeneracies (mass vs.~\(e\)) is essential; otherwise a nondetection could reflect an underestimated tidal-power prior, not a true interior constraint.

\subsection{L 98-59 d Case Study} \label{sec:casestudy}
L98-59d is an OZ planet which been highlighted as a best-case scenario for detecting exoplanet outgassing due to its high nominal eccentricity and favorable transit spectroscopy metric \citep{seligman2023}, and transmission spectroscopy from JWST hints at atmospheric sulfur which may be indicative of volcanic outgassing \citep{Gressier_2024, Banerjee_2024}. Here, we evaluate L98-59d against the framework we have developed in this paper (Sections~\ref{sec:tidal}, \ref{sec:cond}). 

\paragraph{Condition 1: Adequate $e$}
\citet{Cadieux2025} report  $e=0.006^{+0.007}_{-0.004}$ and $P=7.45$ days for L 98-59d. These parameters are compatible with strong tidal heating, with the nominal values corresponding to 1.3$\times$ 10$^{3}$ times Earth's heat flux, competetive with escape (Table~\ref{tab:life}). This constraint is additionally backed up by the 2:1 mean-motion resonance between L~98-59~c and d \citep{Cadieux2025}.
\paragraph{Condition 2: Adequate $m_{H_{2}O}$}
We estimate \(m_{\ch{H2O},\min}\geq1\%\) to sustain a geologically long-lived H$_2$-dominated atmosphere on L 98-59d (Fgiure \ref{fig:inventime}). While L 98-59d's $m_{H_{2}O}$ cannot be measured directly, its low bulk density is compatible with a substantial volatile component \citep{Cadieux2025}. Recent volatile-evolution modeling finds that formation with \(<10{,}000\)~ppmw hydrogen is unlikely \citep{nichollas2025b}, corresponding to \(m_{\ch{H2O}}\geq 8\%\), consistent with this requirement.
\paragraph{Condition 3:Presence of BMO}
Similar to the volatile inventory, a BMO is not directly observable and must be inferred from interior structure and thermal evolution models constrained by measured planetary properties. Recent work reports that the bulk density of L 98-59d favors a substantially molten mantle ($\sim 45\%$), and the threshold $e>0.007$ required to sustain such a mantle is allowed within $2-\sigma$ by updated observations \citep{nichollas2025a,nichollas2025b, Cadieux2025}. This is consistent with our requirement of a substantial mantle magma reservoir, though the modeling does not rule on whether this reservoir is basal or surficial.  

\paragraph{Condition 4: \textit{f}O$_2\lesssim$FMQ-4.4}
L 98-59d is outside the HZ, meaning that the \textit{f}O$_2\lesssim$FMQ-4.4 condition is required to maintain an H$_2$-dominated atmosphere over geologic time. No strong constraint on \textit{f}O$_2$ is available, though the nominal $\mu$ inferred from observations has been suggested to indicate reducing conditions ($<$ FMQ-4; \citealt{nichollas2025b}). Additionally, the tentative inference of $\frac{r_{H_{2}S}}{r_{SO_{2}}}>1$ are consistent with a reducing environment, though we emphasize that these detections are not statistically significant and remain tentative hints. Despite the large uncertainties, we can at least state that the data are consistent with condition 4. 

\subsubsection{Does L 98-59d Have a Thin \ch{H2}-dominated Atmosphere?}
As L 98-59d is a good candidate for hosting a thin H$_2$-dominated atmosphere, we next consider whether the available data are compatible with this scenario according to the methodology described in Section~\ref{sec:falsepos}. Fortuitously, \citet{Gressier_2024} and \citet{Banerjee_2024} have measured $\mu \sim 10$ amu for this atmosphere with JWST, implying that it is H$_2$-dominated. Note that despite the large uncertainties on the individual molecular abundances, the quantity $\mu$ is better constrained, as anticipated \citep{Benneke_2012, niraula_2025}. While further observations are required to confirm the atmospheric detection and measurements of \citet{Gressier_2024} and \citet{Banerjee_2024}, it appears a reasonable working hypothesis that L 98-59d has an H$_2$-dominated atmosphere.

\begin{figure}
    \centering
    \includegraphics[scale = 0.4]{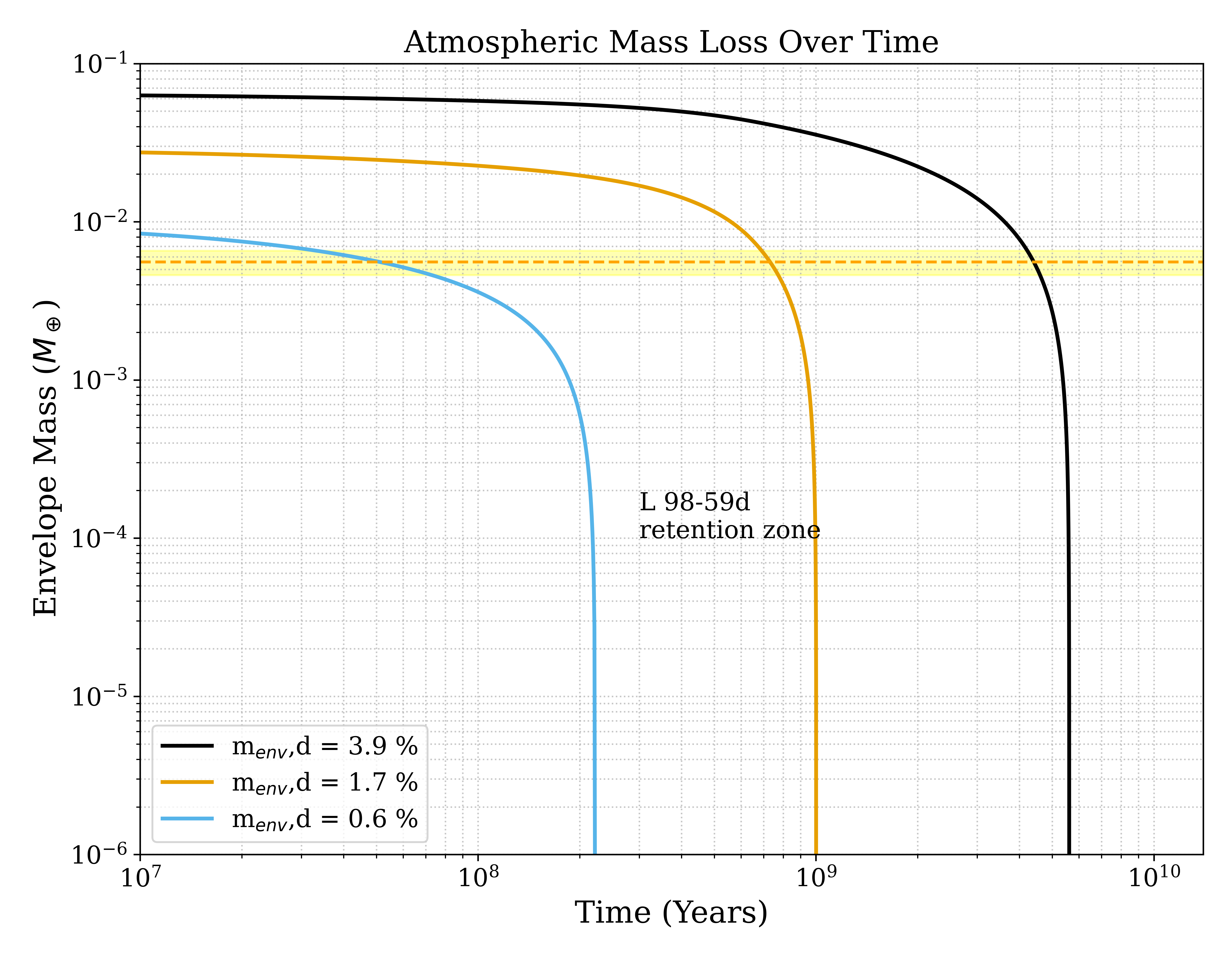}
    \caption{Evolution of envelope mass fraction ($M_{\rm env}/M_\oplus$) over time for L 98-59d, modeled using energy-limited hydrodynamic escape driven by stellar XUV radiation. The shaded regions mark the observationally inferred envelope mass fractions—interpreted as retention zones for the  planet. \citep{Cadieux2025} }
    \label{fig:masslossev_l98}
\end{figure}

Next, we consider whether L 98-59d's atmosphere is thin, and therefore signposts current volcanic outgassing, or is thick, and therefore possibly primordial. To do this, we apply the methodology described in Section~\ref{sec:falsepos}. We begin by estimating the accretion and mass-loss history for L 98-59d (Figure~\ref{fig:masslossev_l98}). We find that a primodial atmosphere is allowed for L 98-59d (assuming gas-poor, dust-free accretion). Therefore, we next estimate $P_{100\text{Myr}}$ = 17 bar, corresponding to an envelope mass fraction $f_{\mathrm{env}, 100~\text{Myr}} = 3.8 \times 10^{-5}$. Thus, f$_{\mathrm{env}}\leq3.8 \times 10^{-5}$ would indicate a thin, outgassed atmosphere.
\citet{Demangeon_2021} reported $\log_{10}(f_{\mathrm{env}}) = -6.43^{+2.61}_{-3.27}$ for L 98-59d, favoring a thin atmosphere but consistent with both thick and thin atmospheres within the uncertainties.  Fortuitously, \citet{Cadieux2025} recently report more precise measurements of the mass and radius of L98-59d (R$_{P}$ = 1.627 $\pm$ 0.041 R$_{\oplus}$, M$_{P}$ = 1.64 $\pm$ 0.07 R$_{\oplus}$ and S = 4.97 $\pm$ 0.52 S$_{\oplus}$). We use \texttt{smint} \citep{Piaulet2021} with the planetary structure models of \cite{Lopez_2014} to estimate f$_{\mathrm{env}}$ of L~98-59~d, assuming an Earth-like core overlaid by an H$_2$/He envelope. We use Gaussian priors on the planet’s bulk properties—mass \(M_p = 1.64 \pm 0.07\,M_\oplus\), radius \(R_p = 1.621 \pm 0.041\,R_\oplus\) \citep{Cadieux2025} and age (\(4.94 \pm 1.44\,\mathrm{Gyr}\)) \citep{Rajpaul2024}. For irradiation, we place a Gaussian prior on the incident flux \(F_{\rm irr}\) =4.97$\pm$ 0.52 (rather than on \(T_{\rm irr}\)), computed via Monte Carlo sampling of the stellar effective temperature and radius together with the orbital distance. We then explore the interior/composition parameter space, including the H/He envelope fraction \(f_{\rm env}\), without imposing a logarithmic prior on \(f_{\rm env}\). The posterior is sampled with an affine-invariant MCMC using \texttt{emcee}; we run 200 walkers for 10000 steps, discard the first 60\% as burn-in, and use the remaining samples to compute the final posterior distributions \citep{Piaulet-Ghorayeb_2024}.
We find f$_{\mathrm{env}}$ = 3.4 $\times$ 10$^{-3}$ $\pm$ 6 $\times$ 10$^{-4}$, which rules out f$_{\mathrm{env}}\leq3.8 \times 10^{-5}$ at 5.6$\sigma$. While future work should repeat this analysis using the three-component model of \citet{Demangeon_2021} instead of the two-component model of \texttt{smint}, it appears most likely to us that L 98-59d has a thick atmosphere of ambiguous origin, either primordial or outgassed.

In summary, L 98-59d, while a promising candidate, does not host a thin H$_2$-dominated atmosphere. Its thick H$_2$-dominated atmosphere is consistent with either a primordial or an outgassed origin \citep{Cadieux2025, nichollas2025b}, and further theoretical and observational work is required to disentangle its nature and origins. However, this worked example of L98-59d illustrates how thin H$_2$-dominated atmospheres can be used to check for volcanic outgassing at low observational cost. Notably, no separate observations had to be obtained for this test -- extant atmospheric surveys and planetary characterization standardly executed for observationally favorable small exoplanets were leveraged, enabling a low cost/high risk/high reward diagnostic. 

\subsection{Are non-detections informative? (L~98-59c Case Study)}
\noindent JWST observations of rocky exoplanets have frequently yielded atmospheric non-detections, consistent with high $\mu$ or airless worlds \citep{Kirk_2024,Radica_2025}. These outcomes are not necessarily null results: for eccentric planets with high tidal heat flux, they can impose meaningful constraints on volatile inventories, mantle redox state, and outgassing efficiency (Section~\ref{subsec:nondetections_eccentric}). We illustrate this with the case study of L 98-59 c.

L~98-59~c is a candidate OZ planet with a non-detection of a \ch{H2}-dominated atmosphere: JWST observations rule out atmospheric metallicities $\lesssim 300\times$ solar ($\mu \gtrsim 10$~amu) at $3\sigma$ confidence \citep{scarsdale_2024}. The planet is likely eccentric ($e=0.002^{+0.002}_{-0.001}$, near-resonance; \citealt{Cadieux2025}), meaning that this nondetection implies that L 98-59 c does not host a reducing water-rich magmatic interior (in potential contrast to L 98-59d; \citealt{nichollas2025b}). 

The principal interpretive challenge is degeneracy: insufficient outgassing, low volatile inventory, or an oxidizing mantle can all yield the same observational outcome. Under our tidal-heating model, a water-rich BMO ($m_{\ch{H2O}} = 3.3\times10^{-2}$) would permit an H$_2$-dominated atmosphere for $1.6\times10^{-2} \lesssim \eta_{\mathrm{og}}\eta_{\mathrm{Heat}} \lesssim 3\times10^{-1}$ (Fig.~\ref{fig:inventime}), and the nondetection of such an atmosphere suggests $\eta_{\mathrm{og}}\eta_{\mathrm{Heat}}$ is outside this range. Alternatively, a mantle more oxidized than FMQ$-3$ would favor \ch{H2O}-dominated outgassing. Thus, a future detection of a \ch{H2O}-dominated atmosphere would point toward oxidizing interior conditions, whereas a deeper atmospheric non-detection would instead suggest that either outgassing efficiency or the volatile budget is insufficient to sustain a detectable secondary atmosphere. Retrievals are required to more precisely quantify these constraints (Section~\ref{sec:caveats}).

More broadly, non-detections should be treated as soft constraints that help shape our understanding of exoplanet populations: even without positive spectral features, they exclude long-lived, low-$\mu$ atmospheres in the OZ and steer interpretations toward oxidized interiors, limited volatile budgets, or inefficient outgassing.

\subsection{Future Targets for \ch{H2} Atmosphere Detection}
\begin{table*}[t]
\centering
\caption{Potential targets for detecting volcanically sustained \ch{H2} atmospheres.}
\label{tab:targets}
\begin{tabular}{lccccccccc}
\hline
Planet & $M_p$ ($M_\oplus$) & $R_p$ ($R_\oplus$) & $e$ & $P$ (days) & $a$ (au)  & $M_\star$ ($M_\odot$) & Age (Gyr) & $d$ (pc) & TSM \\
\hline
HD 260655 b & 2.14 & 1.24 & 0.039 & 2.77 & 0.0293 & 0.439 & 5.0 & 10.0 & 151.6 \\
L 98-59 c   & 2.00 & 1.33 & 0.002 & 3.69 & 0.0309 & 0.292 & 4.9 & 10.6 & 160.7 \\
LHS 1140 c  & 1.91 & 1.27 & 0.050 & 3.78 & 0.0270 & 0.184 & 5.0 & 15.0 & 116.6 \\
\hline
\end{tabular}
\end{table*}
In Table~\ref{tab:targets}, we summarize four nearby rocky planets that emerge as promising candidates for searching for volcanic  \ch{H2} atmospheres based on their nominal planetary parameters. These targets were selected using the following criteria: (i) $R_p < 1.6 \, R_\oplus$, favoring rocky compositions \citep{Rogers2015}; (ii) maximum estimated outgassing rates greater than the expected XUV-driven \ch{H2} escape rate, so that secondary atmospheres can be sustained; (iii) distances within 100~pc, favoring observational feasibility; and (iv) a transmission spectrum metric (TSM) above 100 \citep{Kempton_2018}. Together, these filters identify systems that are both physically favorable and observationally accessible.  


HD~260655~b and LHS~1140~c satisfy the conditions described above for hosting volcanically replenished \ch{H2} envelopes. Yet, in both cases the present uncertainties in mass and radius prevent robust inference of atmospheric composition. Improved constraints on their bulk densities are therefore essential.

L~98-59~c stands out as the most promising near-term target. While two JWST transit observations have ruled out $\mu \lesssim 10$ amu at $3\sigma$ \citep{scarsdale_2024}, higher-$\mu$ mixtures dominated by \ch{H2} with significant \ch{N2} or \ch{CO2} remain consistent with the data. Additional observations are needed to confirm or rule out the presence of such hybrid \ch{H2}-dominated atmospheres. A positive detection would directly link interior outgassing to observable atmospheric composition, while continued non-detections would provide constraints on the volatile budget and redox state of the mantle.  

\subsection{Limitations and Future Work \label{sec:caveats}}
Here we discuss the limitations of our analysis, which are also targets for future work. The largest limitation of our analysis is that our interior physics modeling is simplified. We calculate tidal heating using the runaway-melting prescription of \citet{seligman2023}, whereas in reality the tidal heat flux may be much smaller due to decreased frictive dissipation from a molten mantle \citep{nichollas2025a}. We currently account for this uncertainty by via a parametric heating efficiency $\eta_{\rm Heat}$; a better approach would be to use a detailed interior model, similar to \citet{nichollas2025a}, and ideally rule between the partial melt and runaway melting debate in the literature \citep{seligman2023, nichollas2025a}. Similarly, we currently treat the presence of a basal magma ocean as a free parameter; an improved approach would be to calculate the mantle melt fraction, similar to \citet{nichollas2025a}. If we could use models to constrain a basal magma ocean, it would remove one whole axis of the current degeneracy in interpreting atmospheric nondetections for OZ exoplanets. Next, we follow previous work in assuming melt production scales linearly with heat flux \citep{Hu2023, Gkouvelis2025}. 
However,  melt production may not scale linearly with heat flux, especially at high melt fractions \citep{Sleep2001, Veenstra2024,nichollas2025a}. Future work should aim to more accurately estimate melt production on eccentric exoplanets. Finally, our solubility and partitioning relations are formally valid only over limited temperature–pressure-composition ranges. Following previous work, we extrapolate these laws when needed to cover our large parameter space, which we justify on the basis that it successfully reproduces solar system outgassing \citep{gailard_framework,liggins1,Wogan_2020}. However, while such extrapolations are valid in the Solar System it is conceivable that they may fail on exoplanets. We therefore advocate for measurements of melt solubility laws and partitioning relations over a broader range of conditions. 

We have shown that for eccentric OZ planets, lower limits on $\mu$ from JWST atmospheric nondetections can be used to show absence of a water-rich basal magma ocean, inefficient outgassing, or high \textit{f}O$_2$. However, this constraint is degenerate and qualitative. To translate this qualitative constraint into quantitative limits on \textit{f}O$_2$, $m_{\ch{H2O}}$, and $\eta_{og}\eta_{Heat}$, a retrieval approach must be implemented. To implement a retrieval framework to interpret a $\mu$ lower limit for a given planet, a grid of forward models for bulk specific planet spanning fO$_2$, $m_{\ch{H2O}}$, and $\eta_{og}\eta_{Heat}$ must be built (e.g., \citealt{nichollas2025b} for L 98-59 d), which can be used to build up posteriors for these parameters by MCMC comparison to the $\mu$ constraint (e.g., \citealt{Piaulet2021}). Importantly, such a framework can incorporate the effects of the (substantial) uncertainties on planetary parameters like $e$. The parameter space can be de-dimensionalized if $m_{\ch{H2O}}$ or $\eta_{og}\eta_{Heat}$ can be constrained from interior modeling. Such constraints can be used to determine how common are volcanogenic false positives for reducing gases as biosignatures \citep{Seager2025}, with direct implications for design of and interpretation of data from future missions like the Habitable Worlds Observatory \citep{Clery2023}. We advocate for such work. 



\section{Conclusion}
We propose that (\(\mathrm{H_2}\)-dominated) atmospheres are possible on eccentric terrestrial exoplanets closely orbiting their host stars. Such atmospheres are generally unstable to escape \citep{liggins1, Hu2023}, but we argue that planets subject to high tidal heating can outgas enough H$_2$ to counterbalance escape. Such atmospheres can only be expected over a narrow range of parameter space: in addition to high eccentricities, long-lived H$_2$-dominated atmospheres on terrestrial planets require $m_{\ch{H2O}}\gtrsim1$ \%, a basal magma ocean, and reduced melts (e.g., \(\textit{f}\ch{O2}\lesssim\) FMQ-4.4 for L~98-59~d), which both enhance \(\mathrm{H_2}\) outgassing and suppress heavy gases. A specifically thin H$_2$-dominated atmosphere ($P\leq P_{steady}\sim10$ bar) is a sign of current magmatic volcanic outgassing, because the lifetime of such an atmosphere to escape is geologically short. The main false positive for this outgassing signature, primordial nebular atmospheres, can be discriminated by high-precision mass-radius measurements. As a corollary, we suggest that mini-Neptune atmospheres are not necessarily primordial; eccentric terrestrial exoplanets with water-rich reducing interiors can evolve thick H$_2$-dominated atmospheres.

We delineate a planet-specific outgassing zone (OZ), where H$_2$-dominated atmospheres are most likely to exist. Counterintuitively, this OZ orbits close to the star, due to the stronger inverse dependence of tidal heating on semimajor axis compared to escape. To search for H$_2$-dominated atmospheres, we suggest using radial velocities and transits to precisely constrain mass, radius and eccentricity to confirm OZ candidates, followed by JWST observations to search for H$_2$-dominated atmospheres. Such atmospheres can be detected in a few transits with JWST, making them highly observationally tractable. Detecting H$_2$-dominated terrestrial planet atmospheres offers a probe of interior conditions: a low-\(\mu\) detection in the OZ implies ongoing volcanism and a reducing, water rich interior, whereas a robust non-detection constrains melt redox, mantle melt fraction and volatile inventory, and tidal heating and outgassing efficiency. In particular, systematic nondetections of H$_2$-dominated atmospheres on OZ planets may constrain volcanogenic false positives for reducing biosignature gases to be unlikely, with direct application to biosignature search; further work is required to quantify this possibility. In addition, atmospheric probes of a broad range of surface processes, from geology to biology, would be much more detectable on such objects. We catalog favorable targets to search for H$_2$-dominated atmospheres. 

\section*{Acknowledgment} We thank Laura Schaefer for insightful conversations. R. A. and S. R thank the Kavli Foundation for support (SciAlog grant \#PS-2023-GR-29-2811). R. A., S. R., P. M, and A. M. thank the National Science Foundation (AAG/GLOW Grant \#2408747) for support. This research has made use of the This research has made use of NASA’s Astrophysics Data System. Code associated with this paper is available at \citet{rahul2013396_2025_17238676}

\appendix
\section{Parameter space and physical motivation}

Our outgassing model contains 5 key parameters which must be prescribed ($P_{tot}$, $m^{\mathrm{tot}}_{\mathrm{CO_{2}}}, m^{\mathrm{tot}}_{\mathrm{H_{2}O}}, m^{\mathrm{tot}}_{\mathrm{S}}$, $f\mathrm{O}_{2}$; e.g., \citealt{Wogan_2020}). In Table~\ref{tab:parameterraneg}, we describe and justify the parameter space we explore with our outgassing model. 

\begin{table}[h!]
\centering
\caption{Parameter Ranges for Outgassing Model. After \citet{Wogan_2020}.\label{tab:parameterraneg}}
\begin{tabular}{lcp{9cm}}
\hline 
\textbf{Variable} & \textbf{Range} & \textbf{Justification}\\
\hline
$P_{tot}$ & $10^{-3}$--$10^{3}$ bar & Rough range of subaerial degassing pressures in the solar system ($10^{-3}$--100 bar). The upper limit is obtained from degassing pressure at 1--10 km ocean depth ($\sim$1000 bar) \citep{Wogan_2020}.\\
$m^\mathrm{tot}_\mathrm{CO_{2}}$ & $10^{-5}$--$10^{-2}$ & Approximate CO$_2$ mass fraction range in terrestrial magmas \citep{WALLACE2015163,anderson2017,leVoyer2019}.\\
$m^\mathrm{tot}_\mathrm{H_{2}O}$ & $10^{-5}$--$10^{-1}$ & Range of H$_2$O concentrations in Earth magmas from outgassing \citep{WALLACE2015163}.\\
$m^\mathrm{tot}_\mathrm{S}$ & $10^{-4}$--$10^{-3}$ & Observed in Earth samples of a volcanic rock with a composition between trachyte and basalt \citep{Taracsak2023}.\\
$f\mathrm{O}_2$ & $\mathrm{FMQ}-8$ to $\mathrm{FMQ}+5$ & Oxygen fugacity range from most reducing Martian meteorites to highly oxidized terrestrial magmas \citep{stamper2014oxidation,2017aeil.book.....C}.\\
\hline
\end{tabular}
\end{table}

\section{Solubility Laws of \ch{H2O} and \ch{CO2} in Magma}
\label{sec:appsolub}
Here we discuss the details to calculate the solubility of \ch{H2O} and \ch{CO2}. The solubility of volatile species in silicate melts depends on pressure, temperature, and the chemical composition of the melt. For anhydrous cases, the solubility of \ch{H2O} and \ch{CO2} is modeled following the formalism of \cite{IACONOMARZIANO20121}. 

The solubility of water is given by:

\[
S_{\ch{H2O}} = \ln\left( \frac{\mu_{\mathrm{magma}}}{\mu_{\ch{H2O}} \cdot 10^2} \right) + \frac{C_{\ch{H2O}} P}{T} + B_{\ch{H2O}} + b_{\ch{H2O}} \left[ \frac{\mathrm{NBO}}{O} \right]
\]

Similarly, the solubility of carbon dioxide is expressed using analogous parameters:

\begin{align}    
S_{\ch{CO2}} = \ln\!\left(\frac{\mu_{\mathrm{magma}}}{\mu_{\mathrm{CO_2}}\,10^{6}}\right)
+ \frac{C_{\mathrm{CO_2}}\,P}{T}
+ B_{\mathrm{CO_2}}
+ b_{\mathrm{CO_2}}\!\left[\frac{\mathrm{NBO}}{\mathrm{O}}\right]
+ \left(\frac{x_{\ch{Al2O3}}}{x_{\ch{CaO}} + x_{\ch{K2O}} + x_{\ch{Na2O}}}\right)
  d_{\ch{Al2O3}/(\ch{CaO}+\ch{K2O}+\ch{Na2O})}
+ \\ (x_{\ch{FeO}} + x_{\ch{MgO}})\, d_{\ch{FeO}+\ch{MgO}}
+ (x_{\ch{Na2O}} + x_{\ch{K2O}})\, d_{\ch{Na2O}+\ch{K2O}} \
\end{align}

Here \( \mu \) is the molecular weight,\( T \) is temperature (in K),\( P \) is degassing pressure (in bars),\( \left[ \frac{\mathrm{NBO}}{O} \right] \) is the non-bridging oxygen per oxygen atom in the melt, calculated as:

\[
\left[ \frac{\mathrm{NBO}}{O} \right] = \frac{2(x_{\ch{K2O}} + x_{\ch{Na2O}} + x_{\ch{CaO}} + x_{\ch{MgO}} + x_{\ch{FeO}} - x_{\ch{Al2O3}})}{2x_{\ch{SiO2}} + 2x_{\ch{TiO2}} + 3x_{\ch{Al2O3}} + x_{\ch{MgO}} + x_{\ch{FeO}} + x_{\ch{CaO}} + x_{\ch{Na2O}} + x_{\ch{K2O}}}
\]

The constants \(C, B, b\) are empirically determined and listed in Table 4.\citealt{Wogan_2020} found that different mafic magma compositions do not affect the outputs of the outgassing speciation model therefore, for the purposes of calculating melt solubility, we fixed the chemical make-up of the magma to the magma erupting at Mount Etna as described in \citealt{IACONOMARZIANO20121}. This formulation allows for predicting volatile solubility across a wide range of basaltic melt compositions.

\begin{table}[h!]
\centering
\begin{minipage}{0.48\textwidth}
\centering
\caption{Mount Etna Magma Composition}
\begin{tabular}{@{}lr@{}}
\hline
\textbf{Magma Component} & \textbf{Mole Fraction} \\
\hline
$x_{\ch{SiO2}}$     & 0.516 \\
$x_{\ch{TiO2}}$     & 0.014 \\
$x_{\ch{Al2O3}}$    & 0.110 \\
$x_{\ch{FeO}}$      & 0.091 \\
$x_{\ch{MgO}}$      & 0.092 \\
$x_{\ch{CaO}}$      & 0.126 \\
$x_{\ch{Na2O}}$     & 0.035 \\
$x_{\ch{K2O}}$      & 0.002 \\
$x_{\ch{P2O5}}$     & 0.016 \\
\hline
\end{tabular}
\end{minipage}
\hfill
\begin{minipage}{0.48\textwidth}
\centering
\caption{Solubility Constants \citep{IACONOMARZIANO20121}}
\begin{tabular}{@{}lr@{}}
\hline
\textbf{Constant} & \textbf{Value} \\
\hline
$a_{\ch{H2O}}$ & 0.54 \\
$a_{\ch{CO2}}$ & 1 \\
$s_{\ch{H2O}}$ & 2.3 \\
$C_{\ch{CO2}}$ & 0.14 \\
$B_{\ch{CO2}}$ & $-5.3$ \\
$b_{\ch{CO2}}$ & 15.8 \\
$B_{\ch{H2O}}$ & $-2.95$ \\
$b_{\ch{H2O}}$ & 1.24 \\
$d_{\ch{Al2O3}/(\ch{CaO}+\ch{K2O}+\ch{Na2O})}$ & 3.8 \\
$d_{\ch{FeO}+\ch{MgO}}$ & $-16.3$ \\
$d_{\ch{Na2O}+\ch{K2O}}$ & 20.1 \\
\hline
\end{tabular}
\end{minipage}
\end{table}

\section{Model validation}

\subsection{Outgassing model: comparison with \textsc{VolcGasses}}
\begin{figure}[h!]
    \centering
    \includegraphics[width=\textwidth]{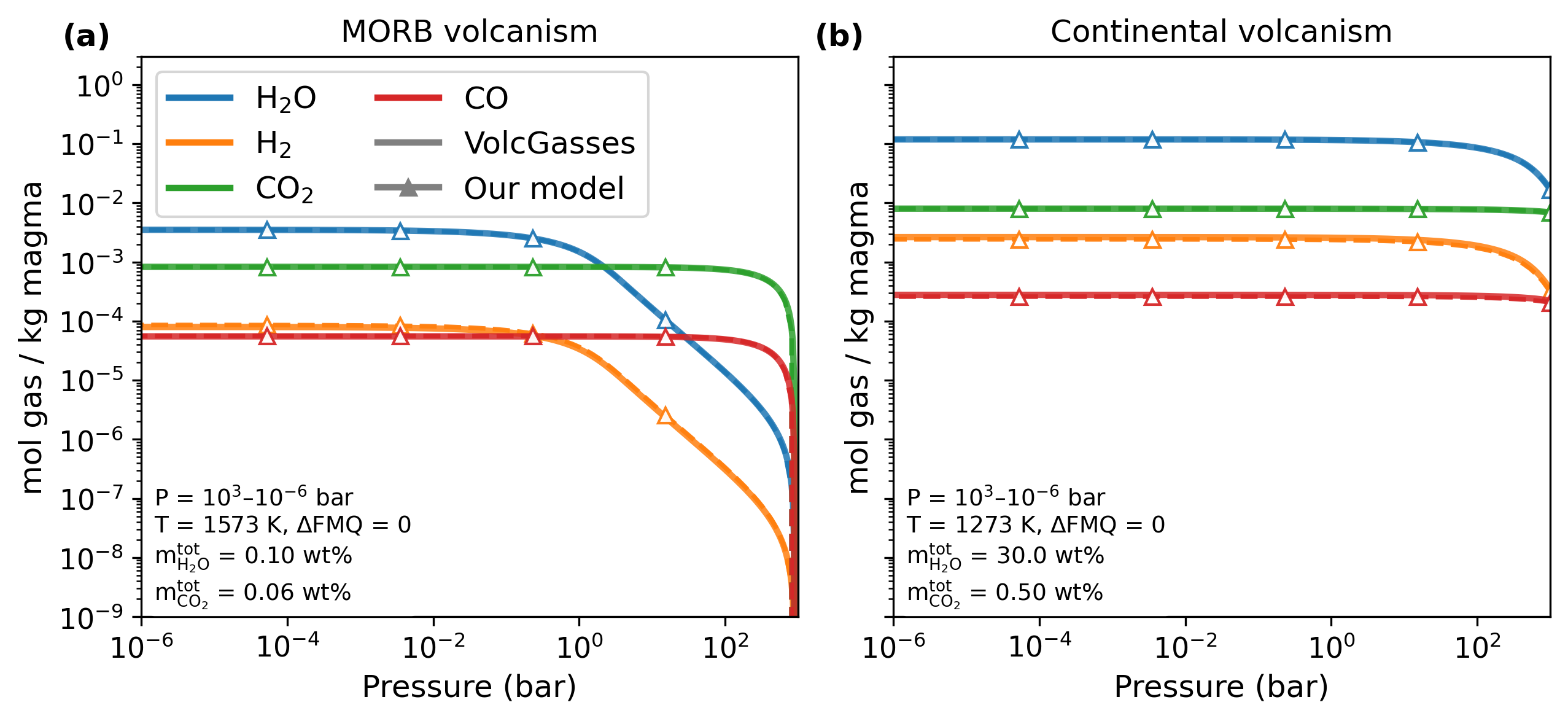}
    \caption{\textbf{Validation of the outgassing module against \textsc{VolcGasses}.}
    Solid colored curves: our model; dashed colored curves: \textsc{VolcGasses}; open triangles: sparse samples along our curves. 
    (a) MORB volcanism: $P=10^{3}$–$10^{-6}$~bar, $T=1573$~K, $\Delta\mathrm{FMQ}=0$, $m_{\mathrm{H_2O}}^{\mathrm{tot}}=0.10$~wt\%, $m_{\mathrm{CO_2}}^{\mathrm{tot}}=0.06$~wt\%. 
    (b) Continental volcanism: same $P$ range, $T=1273$~K, $\Delta\mathrm{FMQ}=0$, $m_{\mathrm{H_2O}}^{\mathrm{tot}}=0.10$~wt\%, $m_{\mathrm{CO_2}}^{\mathrm{tot}}=0.50$~wt\%.
    Axes show mol gas per kg magma on a logarithmic scale.}
    \label{fig:validation_volcgasses}
\end{figure}
We benchmark our equilibrium outgassing module against \textsc{VolcGasses} \citep{Wogan_2020}, using identical bulk volatile inventories, temperature, pressure range, and redox buffer. Figure~\ref{fig:validation_volcgasses} shows two end-member scenarios: (a) MORB-like melts at $T=1573$~K with $m_{\mathrm{H_2O}}^{\mathrm{tot}}=0.10$~wt\% and $m_{\mathrm{CO_2}}^{\mathrm{tot}}=0.06$~wt\% \citep{saal2002,marty2012}, and (b) a CO$_2$-richer continental-style melt at $T=1273$~K with $m_{\mathrm{H_2O}}^{\mathrm{tot}}=3.0$~wt\% and $m_{\mathrm{CO_2}}^{\mathrm{tot}}=0.50$~wt\% \citep{Wallace2005_JVGR_VolatilesArcs,MetrichWallace2008_ChemGeo,Plank2013_4wtH2O}. In both cases we set $\Delta\mathrm{FMQ}=0$ and vary total pressure from $10^{3}$ to $10^{-6}$~bar.

Our curves exactly reproduce the expected phase partitioning (gas and melt) and gas speciation of all molecules (H$_2$O, H$_2$, CO$_2$, CO) across six orders of magnitude in pressure, validating the CHO component of our outgassing calculations. 

\subsection{Graphite saturation: comparison with Ortenzi et al. (2020)}
\begin{figure}[h!]
    \centering
    \includegraphics[width=0.62\textwidth]{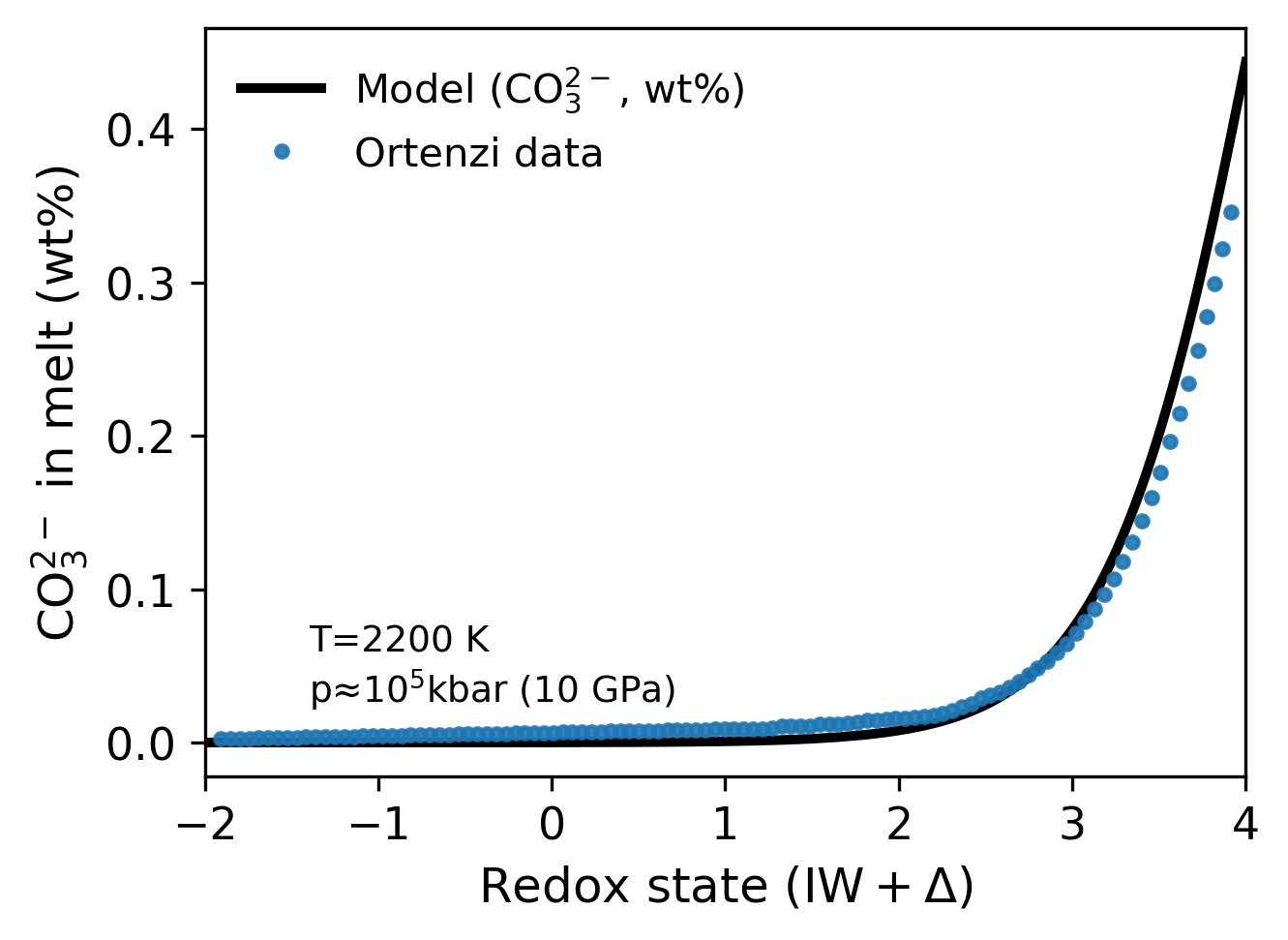}
    \caption{\textbf{Validation of the graphite-saturation / carbonate module.}
    Black line: model prediction of CO$_3^{2-}$ in the melt (wt\%) versus redox state ($\mathrm{IW}+\Delta$) at $T=2200$~K and $p\approx10$~GPa; blue dots: \citet{ortenzi2020}. The model recovers the low-carbonate regime at reducing conditions and the rapid increase at $\mathrm{IW}+\Delta\gtrsim3$.}
    \label{fig:validation_ortenzi}
\end{figure}
To test our treatment of graphite saturation and carbonate formation, we compare the predicted CO$_3^{2-}$ content of the melt to the compilation by \citet{ortenzi2020}. We compute the carbonate fraction from the coupled equilibria parameterized by $K_1(T,P)$ and $K_2(T,P)$, convert mole fractions to wt\% on the same basis as the experiments, and evaluate at $T=2200$~K and $p\approx10$~GPa ($\sim10^5$~bar). The resulting curve (Figure~\ref{fig:validation_ortenzi}) reproduces the near-zero carbonate content under reducing conditions and the sharp increase for $\mathrm{IW}+\Delta \gtrsim 3$, consistent with the transition from CO/graphite-buffered to CO$_2$/carbonate-dominated speciation.

\bibliography{sample631}{}
\bibliographystyle{aasjournal}



\end{document}